\documentclass[sigconf]{acmart}

 \setcopyright{none}

\usepackage{algorithmic}
\usepackage{graphicx}
\usepackage{textcomp}
\usepackage{xcolor}
\usepackage{subcaption} 
\usepackage{enumitem}

\usepackage{graphicx} %

\usepackage{xcolor}
\usepackage{tabularx}
\usepackage{hyperref}
\usepackage{cleveref}

\usepackage{microtype}
\usepackage{etoolbox}

\makeatletter
\apptocmd{\normalsize}{%
  \setlength{\abovedisplayskip}{3pt}%
  \setlength{\belowdisplayskip}{3pt}%
  \setlength{\abovedisplayshortskip}{1pt}%
  \setlength{\belowdisplayshortskip}{1pt}%
}{}{}
\makeatother

\usepackage{array}
\usepackage{makecell}
\usepackage{booktabs}
\usepackage{xcolor}
\usepackage{listings}
\usepackage[most]{tcolorbox}
\usepackage{upquote}

\lstdefinestyle{artifactstyle}{
    language=bash,
    basicstyle=\ttfamily\small,
    commentstyle=\color[HTML]{008000}\itshape,
    columns=fullflexible,
    keepspaces=true,
    breaklines=true,
    breakatwhitespace=false,
    showstringspaces=false,
    upquote=true,
    tabsize=4,
    aboveskip=0pt,
    belowskip=0pt
}

\newtcblisting{artifactbox}{
    listing only,
    listing engine=listings,
    listing options={style=artifactstyle},
    enhanced,
    breakable,
    width=0.96\linewidth,
    center,
    colback=black!4,
    colframe=black!20,
    boxrule=0.4pt,
    arc=1mm,
    boxsep=0pt,
    left=2mm,
    right=1.5mm,
    top=0.5mm,
    bottom=0.5mm,
    before skip=7pt,
    after skip=7pt
}

\lstdefinestyle{artifactresultstyle}{
    style=artifactstyle,
    basicstyle=\ttfamily\footnotesize
}

\newtcblisting{resultbox}{
    listing only,
    listing engine=listings,
    listing options={style=artifactresultstyle},
    enhanced,
    breakable,
    width=0.96\linewidth,
    center,
    colback=black!4,
    colframe=black!20,
    boxrule=0.4pt,
    arc=1mm,
    boxsep=0pt,
    left=2mm,
    right=1.5mm,
    top=0.5mm,
    bottom=0.5mm,
    before skip=7pt,
    after skip=7pt
}
\def\BibTeX{{\rm B\kern-.05em{\sc i\kern-.025em b}\kern-.08em
    T\kern-.1667em\lower.7ex\hbox{E}\kern-.125emX}}

\newcommand{\justin}[1]{\textcolor{orange}{\textbf{Justin: #1}}}

\newcommand{\ignore}[1]{}

\newcommand{\calibration}[1]{}

\newcommand{\shepherd}[1]{#1}

\newcommand{\shepherdtwo}[1]{#1}

\begin{document}

\copyrightyear{2026}\acmYear{2026}
\setcopyright{cc}
\setcctype{by}
\acmConference[PACT '26]{International Conference on Parallel Architectures and Compilation Techniques}{October 19-22, 2026}{Chicago, IL, USA}
\acmBooktitle{International Conference on Parallel Architectures and Compilation Techniques (PACT '26), October 19-22, 2026, Chicago, IL, USA}
\acmDOI{10.1145/3838684.3846832}
\acmISBN{979-8-4007-2915-7/2026/10}

\pdfpagewidth=8.5in
\pdfpageheight=11in

\pagenumbering{arabic}

\newcommand{\name}{Proba}

\title{\name{}: A High-Performance, Low-Traffic Probabilistic Spatial Memory Streaming Prefetcher} %

\author{Yinting Huang}\affiliation{\institution{Huawei Newton Research Centre} \city{Cambridge}\country{United Kingdom}}\email{yinting\_justin\_huang@outlook.com}
\author{Jacky Wong}\affiliation{\institution{Imperial College London} \city{London} \country{United Kingdom}}\email{jacky.wong20@imperial.ac.uk}
\author{Sam Ainsworth}\affiliation{\institution{University of Edinburgh} \city{Edinburgh} \country{United Kingdom}}\email{sam.ainsworth@ed.ac.uk}

\begin{abstract}
Spatial Memory Streaming (SMS) prefetchers exploit recurring spatial access patterns across memory regions to predict future accesses. Despite their effectiveness, existing SMS-based designs suffer from coarse control over aggressiveness and limited adaptability to dynamic program behavior and historical patterns. In this work, we present Proba, a feedback-directed SMS-based prefetcher that integrates probabilistic pattern updates, marginal accuracy decision-making, and pattern accumulation over several trigger features to achieve fine-grained accuracy–coverage trade-offs. When accuracy deteriorates, Proba moderates its aggressiveness to prioritize accurate prefetches; conversely, when accuracy rises, it expands its coverage. Proba achieves higher speedups than any other SMS-based prefetcher while being vastly more efficient in DRAM traffic. This culminates in a large performance gap between Proba and other techniques in multicore and DRAM-constrained scenarios without harming single-threaded throughput.

\end{abstract}


\maketitle

\section{Introduction}
The ever-growing gap between processor performance and memory access latency continues to throttle application performance despite extensive on-chip optimizations. Even though clock frequencies have plateaued since the early 2000s, architectural advances such as out-of-order execution, wider pipelines, and multicore scaling have steadily increased effective processor performance. In contrast, memory latency and bandwidth have improved at a much slower pace. As a result, the gap between processor speed and memory access time has widened, leading to frequent stalls and underutilization of computational resources.

To mitigate this memory wall, processors employ data prefetchers that anticipate future memory accesses and fetch data into on-chip caches before demand arises. One widely deployed prefetcher is Spatial Memory Streaming (SMS)~\cite{SMS,somogyi2011spatial,ferdman2011spatial}. SMS is built on the observation that memory accesses often exhibit spatial regularity: a traversal pattern observed in one memory region is likely to reappear in another. SMS encodes these access patterns in compact bit patterns, indexed by trigger features: \shepherd{metadata associated with the accesses that begin a spatial region, such as the program counter (PC) or the offset of the first access within the region}. When the same trigger feature appears in a different region, SMS replays the prediction stored in the history table. SMS is a lightweight yet effective mechanism for capturing recurring spatial patterns and has been adopted in modern cores such as the Neoverse V2~\cite{Arm-Hotchips}.

\begin{figure}
    \centering
    \includegraphics[width=\columnwidth]{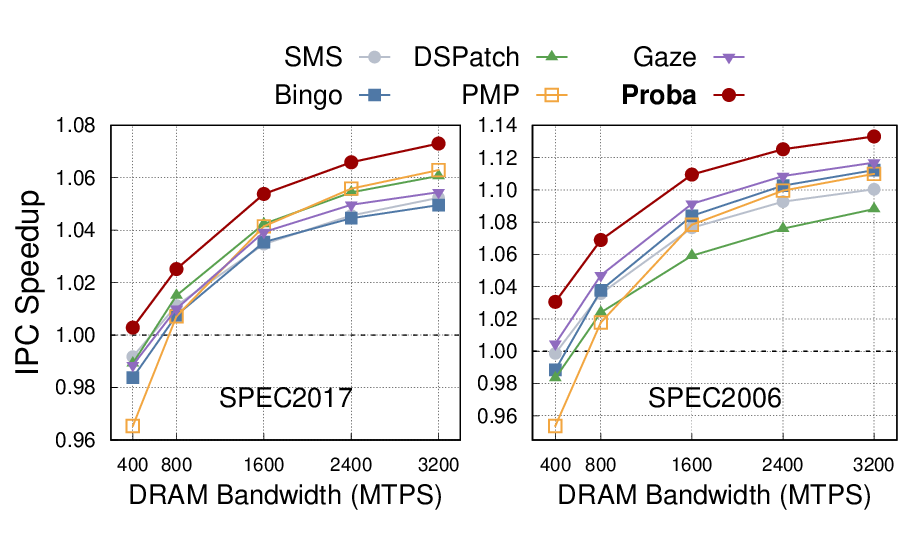}
    \vspace{-20pt}
    \caption{Geomean speedup of SMS-based prefetchers at various DRAM bandwidth constraints. Proba consistently outperforms others because it is highly frugal in traffic.}
    \label{fig:summary-means}
    \vspace{-5pt}
\end{figure}

Although SMS improves performance on many workloads, it incurs substantial memory overhead by issuing prefetches even when access behavior does not exhibit predictive patterns across regions. This excessive memory consumption reduces the efficiency of SMS and can degrade performance under bandwidth-constrained conditions. DSPatch \cite{DSPatch} attempts to address this problem by maintaining two bit patterns: one biased toward coverage and the other biased toward accuracy. At runtime, it selects between these two patterns according to observed DRAM-bandwidth utilization. While effective, this modulation mechanism remains coarse-grained, as it can only modulate between two fixed update behaviors. \ignore{At runtime, DSPatch selects one pattern based on observed DRAM-bandwidth utilization, increasing coverage when headroom exists and throttling to maintain accuracy near peak utilization.} More recent work, PMP \cite{PMP}, improves on this by introducing a pattern-merging strategy based on dual pattern tables and saturating counters. PMP accumulates patterns for different trigger features in separate tables and uses saturating counters to guide which merged pattern should be used for prefetching. However, these counters are storage-intensive and reduce the effective storage capacity available for pattern history, limiting PMP to relatively coarse-grained trigger features and constraining its prediction capability. Furthermore, because both DSPatch and PMP require greater storage per entry, they must compress metadata and coarsen the granularity of pattern updates, which biases them toward coverage at the expense of accuracy.

We propose \name{}, a novel probabilistic update mechanism for SMS-based prefetchers. Instead of interpolating between two history patterns as in DSPatch or accumulating patterns with saturating counters as in PMP, \name{} consolidates alternating patterns into a single, progressively refined prediction, maintaining high accuracy without additional storage overhead. The update probability is dynamically adjusted via accuracy feedback to regulate prefetch aggressiveness. Through gradual probabilistic updates, \name{} adapts smoothly to evolving program access patterns and workload characteristics, improving the performance while mitigating unnecessary DRAM traffic. We further design a \textit{marginal accuracy} mechanism that is able to dynamically adapt predictions using the trigger features of several state-of-the-art SMS-based prefetchers~\cite{DSPatch,SMS,Gaze,PMP}. Finally, we design a Jail Table that decouples pattern discovery from region tracking, enabling precise cache-generation tracking without large storage buffers.

As shown in \Cref{fig:summary-means}, Proba consistently outperforms all state-of-the-art SMS-based prefetchers, regardless of whether DRAM bandwidth is ample or highly constrained. It limits inaccurate prefetches while favoring coverage when additional prefetching is beneficial. This becomes more pronounced in multicore settings, where shared memory bandwidth is scarce. Proba's frugal use of bandwidth makes it well suited to modern energy-bound systems~\cite{EnergyPF,10.1145/3224419} across the scale from mobile phones to data centers.

\section{Background and Motivation}
\subsection{Spatial Memory Streaming (SMS)}
\label{sec:sms_background}

\begin{figure}
    \centering
    \includegraphics[width=0.85\columnwidth]{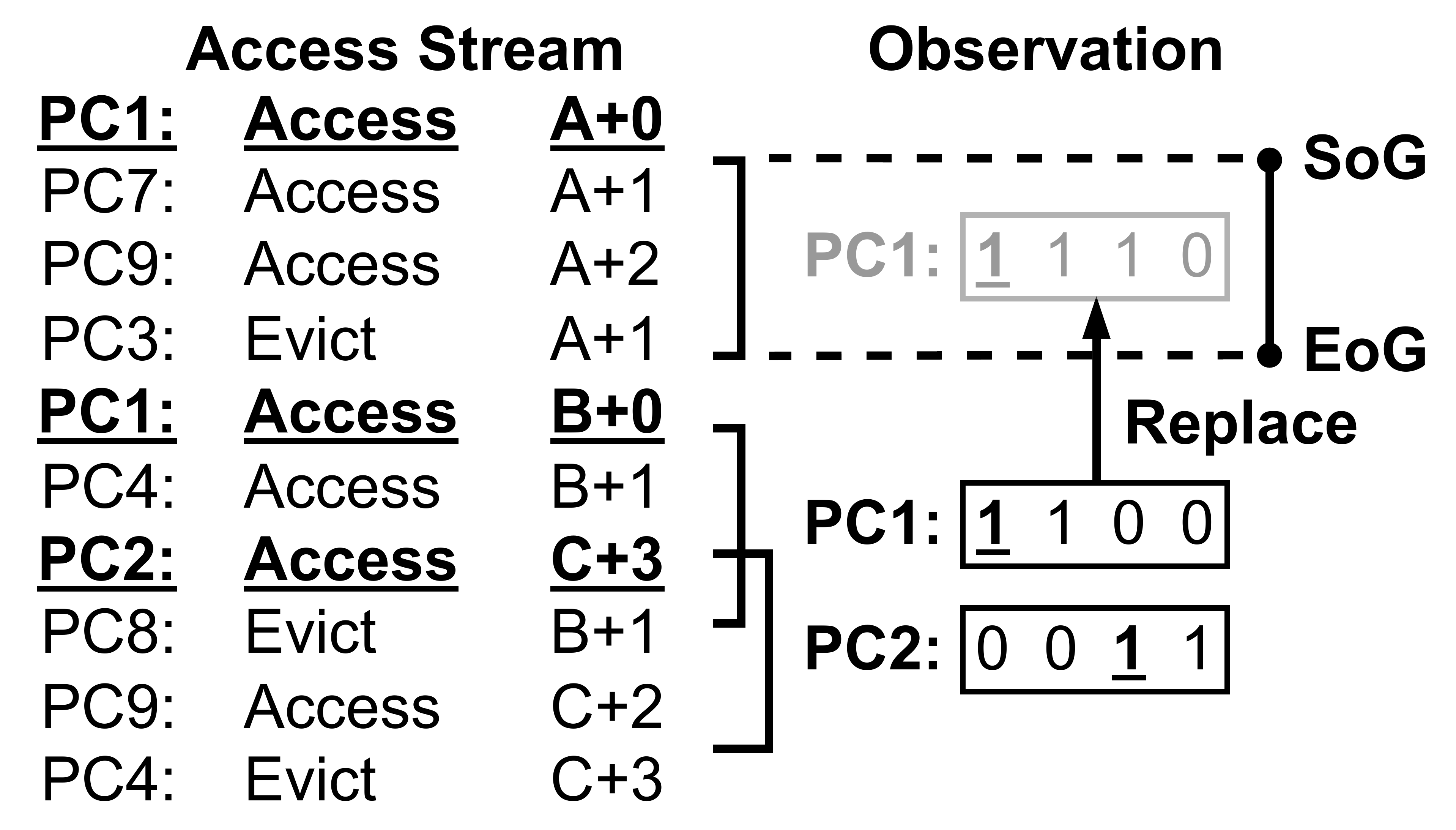}
    \vspace{-10pt}
    \caption{Basic operation of the Spatial Memory Streaming (SMS) prefetcher. SMS records the access pattern observed from the first access to a memory region until its first eviction, and reuses the recorded pattern to predict accesses when the same trigger feature is observed in another region.}
    \label{fig:sms_basics}
    \vspace{-10pt}
\end{figure}

SMS~\cite{SMS} is built on the observation that commercial applications often exhibit recurring spatial patterns across memory regions. To capture these patterns, SMS divides memory into fixed-size spatial regions (e.g., 4KiB). The first access to a region is the trigger access, marking the Start-of-Generation (SoG). For each trigger, SMS records the trigger Program Counter (PC) along with the cache-line access offset in the Active Generation Table (AGT). Subsequent accesses within the same region are recorded by setting the corresponding bit positions in the bit pattern. Tracking of a region terminates when any of the accessed cache lines is evicted from the cache, signaling the End-of-Generation (EoG). The bit pattern is then stored in the Pattern History Table (PHT) indexed by the trigger \texttt{PC+Offset}. When an access with the same trigger feature occurs in a new spatial region, SMS issues the stored bit pattern as prefetches. The basic mechanism of SMS is shown in \Cref{fig:sms_basics}.

\ignore{An extension of SMS~\cite{ferdman2011spatial} observes that spatial patterns associated with the same trigger PC are not unique; they exhibit regularity relative to the trigger offset within a region. To eliminate redundant patterns arising from different offsets with the same trigger PC, they propose rotation-based indexing: before a bit pattern is stored, it is rotated so the trigger access always aligns with the first bit. The rotated pattern is stored in the PHT with a PC-only signature. When issuing prefetches, the PHT is indexed using the trigger \texttt{PC+Offset}, and the original pattern is reconstructed through an inverse rotation based on the trigger offset.} %

\ignore{
Another extension of SMS~\cite{somogyi2011spatial} improves prediction accuracy by incorporating additional spatial history. In the first two variants, instead of storing the most recent pattern, the PHT is expanded to maintain the two most recently observed bit patterns for each PC signature. SMS-OR variant performs a bitwise OR of the two patterns to maximize coverage at the cost of increased overprediction. SMS-AND variant performs a bitwise AND to suppress rarely accessed cache lines and minimize overprediction, though at the risk of reduced coverage. In the third variant, the PHT is expanded to maintain a vector of 2-bit saturating counters, each counter corresponding to a particular cache line in a spatial region. Upon storing a new observation, counters corresponding to cache lines in the observation are initialized to 2, while counters for unaccessed cache lines are initialized to 1. When subsequent observations update an existing PHT entry, SMS increments counters for cache lines that are accessed and decrements those that are not. At prediction time, SMS prefetches all cache lines with counter value greater than 2, biasing prefetch toward consistently observed spatial patterns.

SMS stands out from other address-delta-correlating prefetchers due to its ability to extract irregular but repetitive spatial correlations that do not follow an arithmetic stride or offset. Nevertheless, SMS also suffers from several inherent limitations, as discussed in \Cref{sec:sms_limitation}.
}

\label{sec:sms_limitation}
The baseline SMS lacks the ability to adapt prefetch aggression based on the observed memory accesses. Upon encountering a new access pattern, SMS replaces the existing pattern in the Pattern History Table (PHT) with the new observation, thereby discarding all historical information. This assumes that the most recent observation is more accurate, which is not guaranteed. Consequently, SMS oscillates between patterns and suffers from poor adaptivity. \ignore{Later variants~\cite{somogyi2011spatial} add more history via bitwise AND or OR over the last two updates, or 2-bit saturating counters (also used in more recent designs such as PMP~\cite{PMP}), but this increases storage requirements, and such static policies do not adapt to the level of success actually attainable, where levels of aggression that lead to success are highly workload- and access-dependent.}

\ignore{
SMS-AND, SMS-OR, and SMS-Sat (saturating-counter based) improve upon the baseline SMS design by storing richer spatial history: either by tracking two recent bit patterns and combining them with bitwise operations, or by maintaining a 2-bit counter per cache line. However, all three variants store additional history or metadata per PC signature, which doubles the storage requirement. Besides, SMS-AND and SMS-OR statically support one mode of operation: either extremely accuracy-seeking (bitwise-AND) or extremely coverage-seeking (bitwise-OR), lacking finer grained aggression control. This static design treats all PCs uniformly, disregarding the fact that different PCs may exhibit vastly different behaviors. For instance, some PCs may exhibit poor spatial regularity and benefit from accuracy-seeking to avoid mispredictions, while others with predictable patterns benefit from aggressive prefetching to maximize coverage.
}

\subsection{Dual Spatial Pattern Prefetcher (DSPatch)}
\label{sec:dspatch_background}
\begin{figure}
    \centering
    \includegraphics[width=0.95\columnwidth]{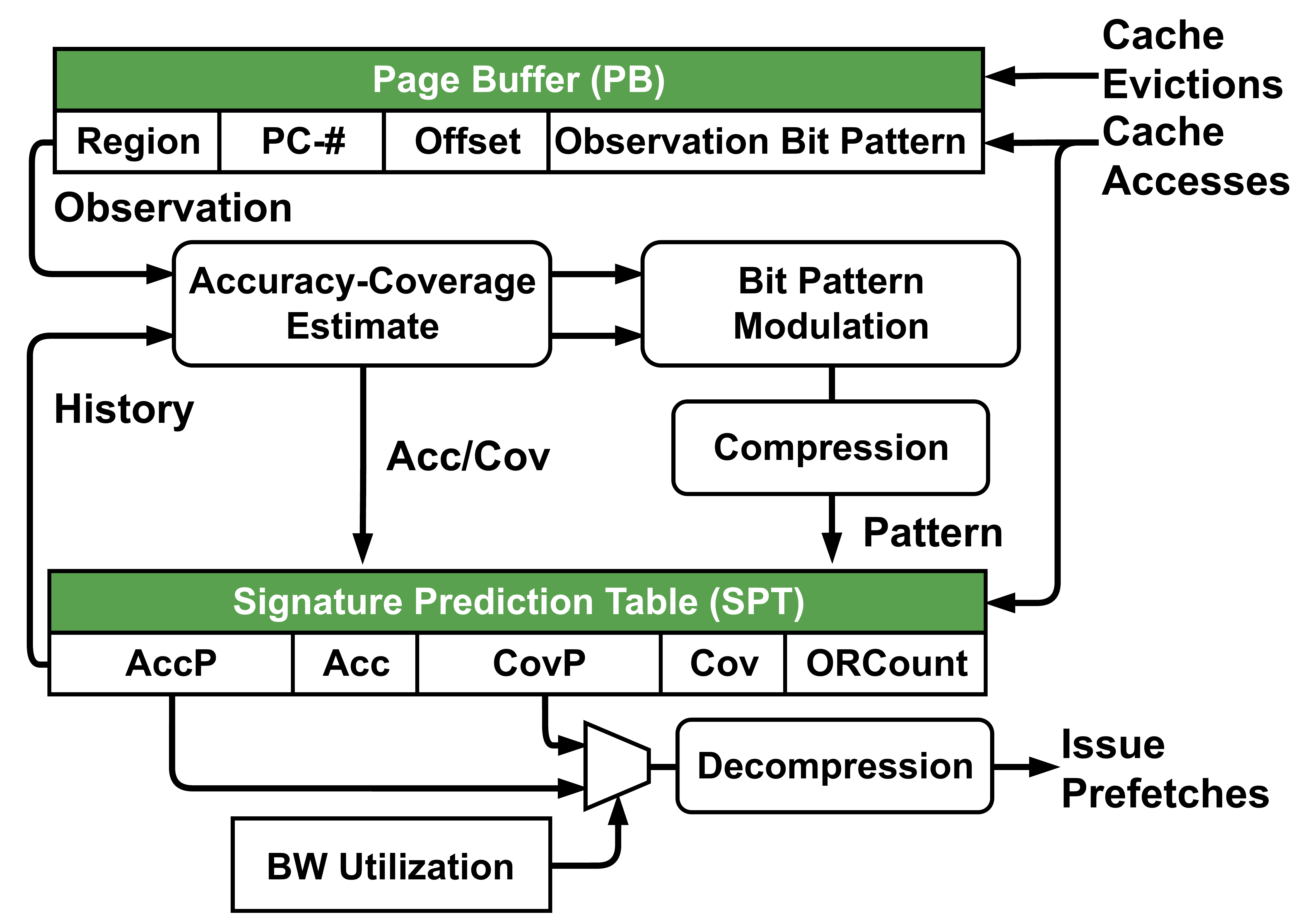}
    \vspace{-10pt}
    \caption{DSPatch tracks active pages in the Page Buffer (PB) and updates their access patterns in the Signature Prediction Table (SPT). SPT entries maintain both a coverage-seeking pattern (CovP) and an accuracy-seeking pattern (AccP). \ignore{DSPatch selects one based on current bandwidth utilization, together with feedback on prefetch accuracy and coverage.}}
    \label{fig:dspatch_basics}
    \vspace{-10pt}
\end{figure}

In response to the limitations of SMS, the Dual Spatial Pattern Prefetcher (DSPatch)~\cite{DSPatch} introduces a mechanism to dynamically balance accuracy and coverage at the pattern level. \shepherd{As shown in \Cref{fig:dspatch_basics}}, DSPatch maintains two modulated bit patterns with associated quantitative metrics: one biased toward accuracy (AccP), the other toward coverage (CovP). DSPatch selects a pattern based on both pattern-level metrics and system bandwidth utilization. When bandwidth utilization is high, DSPatch prioritizes accuracy-biased patterns. When bandwidth utilization is low, it prefetches the coverage-biased pattern if its accuracy is acceptable; otherwise, it issues the accuracy-biased pattern.

DSPatch uses a Page Buffer (PB) to record the trigger PC and offset together with the observed spatial access pattern. Its Signature Prediction Table (SPT) stores two modulated bit patterns per entry. Bit patterns are modulated by combining new observations with stored patterns: bitwise OR operations bias the pattern toward coverage, while bitwise AND operations bias toward accuracy. The SPT is organized as a tagless direct-mapped structure indexed only by the trigger PC, and stores bit patterns anchored to the trigger offset as discussed in the SMS variant~\cite{ferdman2011spatial}. To maintain two bit patterns without doubling the storage requirement, DSPatch compresses them by merging adjacent bits into a single bit and stores these compact representations in the SPT, allowing neighboring cache lines to be encoded compactly. The compression scheme is similar to the one shown in \Cref{fig:pmp_basics}.

\begin{figure}
    \centering
    \includegraphics[width=0.8\columnwidth]{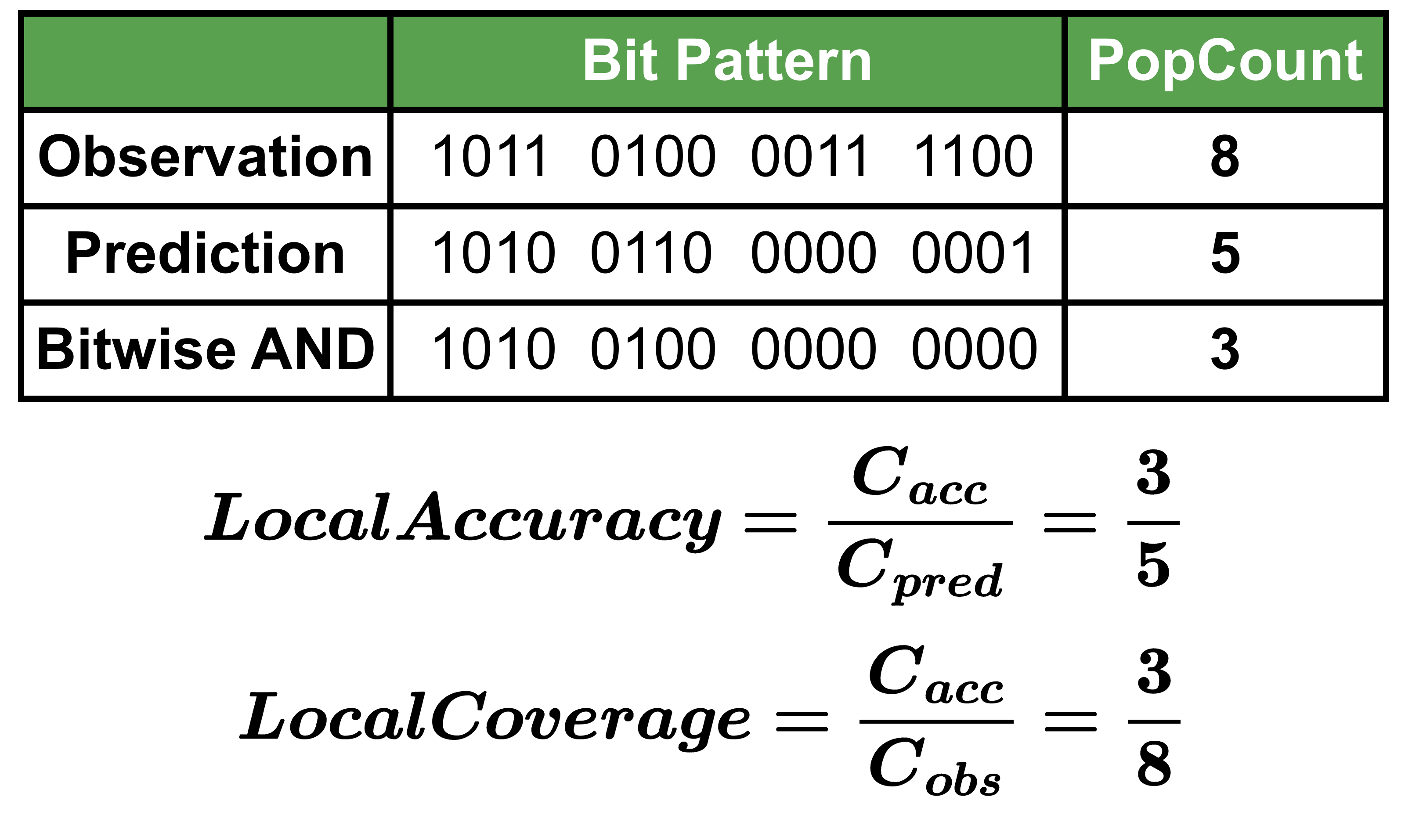}
    \vspace{-18pt}
    \caption{An example of accuracy and coverage estimation used by DSPatch, based on the number of matching bits.}
    \label{fig:dspatch_accuracy_coverage_estimate}
    \vspace{-15pt}
\end{figure}

DSPatch employs two metrics to quantify the accuracy and coverage of its modulated bit patterns at PC granularity. During SPT updates, the popcount of the stored pattern gives the number of predicted prefetches ($C_{pred}$), the popcount of the observed pattern gives the number of actual accesses ($C_{obs}$), and the popcount of their intersection gives the number of accurate prefetches ($C_{acc}$). Accuracy is therefore estimated as $C_{acc}/C_{pred}$, while coverage is estimated as $C_{acc}/C_{obs}$. An example is shown in \Cref{fig:dspatch_accuracy_coverage_estimate}.

To preserve more history than SMS, DSPatch updates its stored patterns through repeated bitwise OR and AND operations. However, this modulation scheme is inherently unstable: repeated OR operations drive the pattern toward all ones, while repeated AND operations drive it toward all zeros. To limit this effect, DSPatch uses a 2-bit saturating counter (\texttt{ORCount}) to track the number of OR updates in each SPT entry. Once the counter saturates, further accumulation is stopped and the pattern is reset and retrained.

Although DSPatch tries to improve the accuracy-coverage trade-off over SMS, its effectiveness is limited by heuristics. First, repetitive OR operations cause the bit patterns to converge toward all ones. Once \texttt{ORCount} saturates, DSPatch requires retraining the entire pattern, resulting in a complete loss of historical information. At the same time, the accuracy-biased pattern is modulated from a coverage-based pattern that becomes increasingly aggressive, and can also inherit this aggressiveness. Compression further exacerbates this problem by biasing predictions toward coverage and introducing substantial inaccuracy. As shown in \Cref{fig:pmp_basics} for PMP~\cite{PMP} where it is also used, pattern reconstruction can introduce prediction errors of up to 50\%. These heuristics make DSPatch highly aggressive in all modes.

\calibration{
Furthermore, we observe that the metrics used by DSPatch often lead to overly optimistic or pessimistic decisions. As shown in \Cref{fig:accuracy_mismatch}, the correctness of the modeled accuracy, in both accuracy-seeking and coverage-seeking modes, varies significantly across workloads. In \texttt{bwaves\_2017} and \texttt{xalan\_2017}, the modeled accuracy closely matches the actual prefetch accuracy. By contrast, for \texttt{astar\_2006} and \texttt{perlbench\_2006}, DSPatch reports high modeled accuracy despite poor actual accuracy. Conversely, in \texttt{zeusmp\_2006}, \texttt{calculix\_2006}, and \texttt{cactu\_2017}, it substantially underestimates accuracy, missing potential gains from higher aggressiveness. 

This imprecise tracking can be attributed to several factors. The Page Buffer (PB) samples access streams but observations are frequently evicted before the full sequence is observed, leading to sampling bias. The estimator also incorrectly classifies prefetch cache hits (which are redundant and harmless as they do not trigger prefetches) as inaccurate, potentially distorting the true measure of prefetch effectiveness. Finally, multiple regions in the PB can be triggered by the same PC, resulting in overlapping updates in the PHT. This introduces discrepancies between patterns used for metric estimation and those used when issuing prefetches.
}

\subsection{Pattern Merging Prefetcher (PMP)}
\label{sec:pmp_background}

\begin{figure}
    \centering
    \includegraphics[width=0.95\columnwidth]{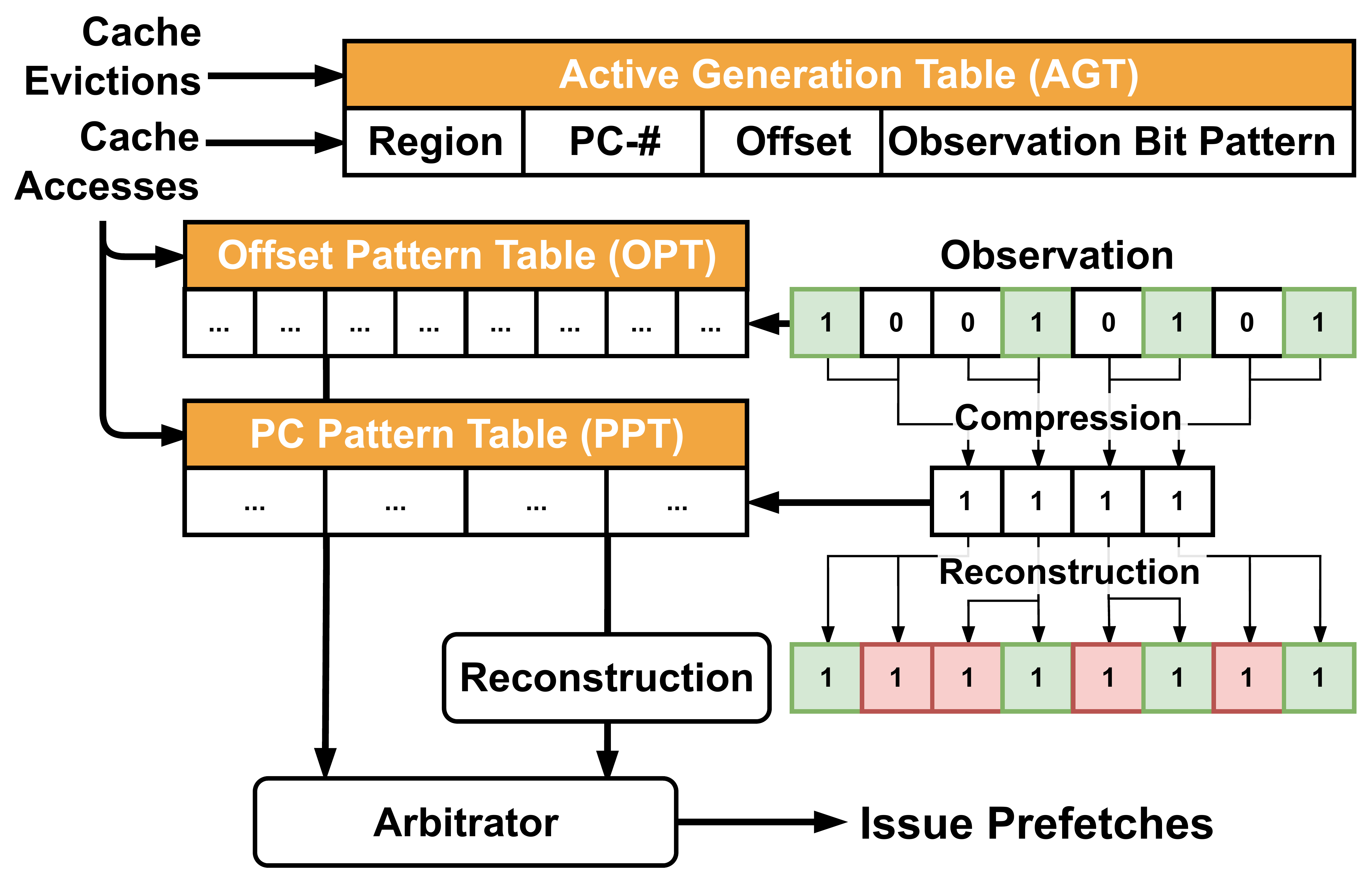}
    \vspace{-10pt}
    \caption{PMP accumulates patterns into the Offset Pattern Table (OPT) and the PC Pattern Table (PPT). The OPT stores the original pattern, while the PPT stores a compressed version. During prediction, PMP combines the OPT prediction with the reconstructed PPT prediction to issue prefetches.}
    \label{fig:pmp_basics}
    \vspace{-10pt}
\end{figure}

The Pattern Merging Prefetcher (PMP)~\cite{PMP} revisits SMS from another direction. PMP (\Cref{fig:pmp_basics}) merges multiple patterns and uses the merged result to guide prefetching. Its key idea is that spatial regions with similar trigger features share structure, which can be exposed by accumulating observations.

Similar to SMS, PMP uses an Active Generation Table (AGT) to detect the start of a generation and record the observed spatial footprint until the region is evicted. Learned patterns are accumulated into two history structures: an Offset Pattern Table (OPT), indexed by the trigger offset (\texttt{Offset-Only}), and a PC Pattern Table (PPT), indexed by the trigger PC (\texttt{PC-Only}). Both tables store counter vectors rather than plain bit patterns; the OPT keeps a fine-grained pattern, while the PPT stores a compressed version of the pattern at a coarser granularity. At prediction time, PMP extracts candidate patterns from both tables and uses arbitration over the counter vectors to determine which offsets should be prefetched and where they should be filled to.

This design improves upon SMS and DSPatch by preserving more than one recent observation and by allowing repeated patterns to reinforce one another over time. However, this comes at a storage cost. Because PMP stores confidence information as counter vectors, a substantial fraction of the hardware budget is spent on metadata instead of pattern capacity. This reduces the effective capacity for distinct histories and forces PMP to rely on coarse trigger features, namely \texttt{PC-Only} and \texttt{Offset-Only}. In contrast, other SMS-based prefetchers show that finer-grained trigger features such as \texttt{PC+Offset} and \texttt{Offset+Offset} can capture important regularity that coarse signatures miss~\cite{ferdman2011spatial,Bingo,Gaze}.

PMP is also limited by the heuristics used to translate merged counters into prefetch decisions. The PPT stores a coarser pattern than the OPT, using DSPatch-like compression, and both tables rely on fixed voting thresholds to make prefetch decisions. Such coarse-grained policies do not adapt to the level of success actually attainable, where the optimal degree of prefetch aggressiveness is highly workload- and access-dependent. Consequently, PMP often generates inaccurate prefetches that increase memory bandwidth consumption, resulting in higher DRAM traffic (\Cref{{fig:speedupdram}}).

\section{Our Solution: \name{}}
\begin{figure*}[t]
    \centering
    \includegraphics[width=\textwidth]{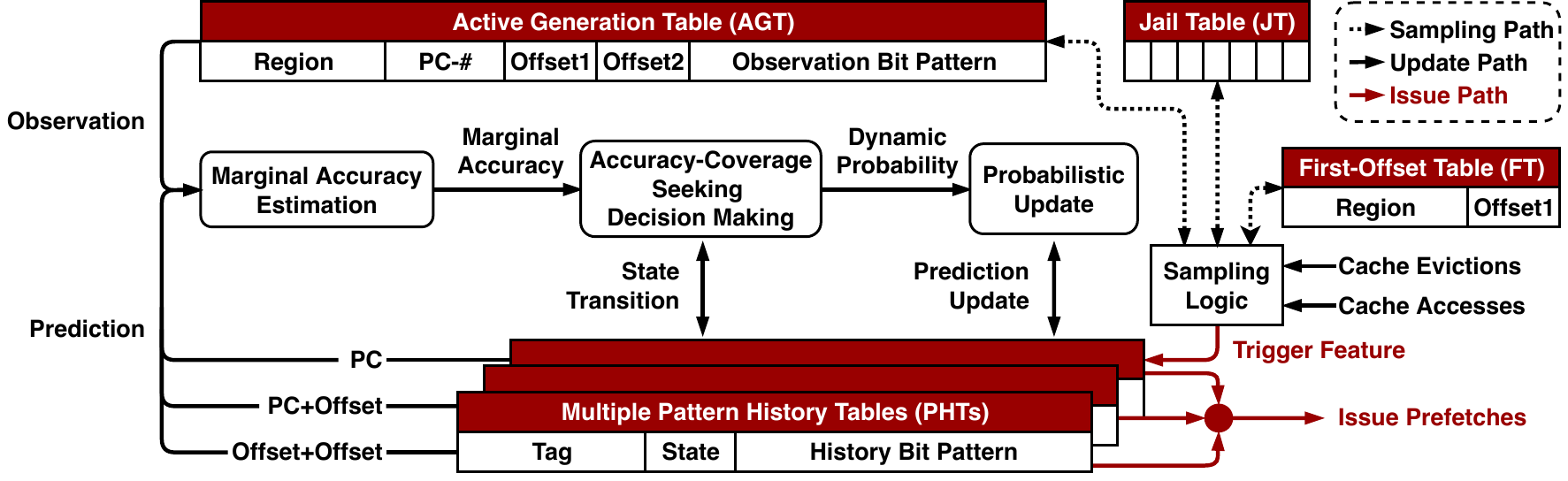}
    \vspace{-20pt}
    \caption{Design overview of Proba. Proba uses probabilistic updates to maintain accurate, high-coverage patterns in its Pattern History Tables (PHTs). Marginal accuracy dynamically controls prefetch aggression, while region sampling and Bloom-like hash-based mappings provide precise observation of cache-region generation behavior.}
    \label{fig:proba}
    \vspace{-5pt}
\end{figure*}
We introduce \textbf{\name{}}, an SMS-based prefetcher that leverages probabilistic updates to maintain useful patterns without requiring extra storage or lossy compression. \name{} treats pattern update as a stochastic process, enabling gradual pattern evolution. \name{} has the following features:
\begin{itemize}[leftmargin=*]
\item \textbf{Probabilistic Update:} Stores a single bit pattern and updates it probabilistically upon each observation, eliminating both the storage overhead and the pattern corruption inherent in prior designs (\Cref{ssec:proba-update}).
\item \textbf{Accuracy-Coverage Seeking Decision-making:} Incorporates accuracy feedback to drive probability state transitions, enabling fine-grained adaptive accuracy–coverage trade-offs based on runtime behavior (\Cref{ssec:seeking}).
\calibration{\item \textbf{Accuracy Calibration:} Refines the PC-local accuracy metric inherited from DSPatch by calibrating it against the global prefetch accuracy, eliminating systematic bias and improving decision precision (\Cref{ssec:calibration}).}
\item \textbf{Multiple Pattern History Tables with Marginal Accuracy Decision-making:} Proba takes the union ~\cite{PMP} of predictions of multiple PHTs indexed by distinct trigger features~\cite{DSPatch,somogyi2011spatial,ferdman2011spatial,Gaze}. Rather than assigning aggressiveness to each table in isolation, Proba adjusts each table according to its \textit{marginal accuracy}: the accuracy of the additional prefetches it contributes beyond those already predicted by more accurate tables (\Cref{ssec:marginalpht}).
\item \textbf{Jail Table (JT) and Region Sampling:} Prevents resource thrashing during pattern learning by decoupling the functionality of the Active Generation Table (AGT) into two parts: a Jail Table, which tracks active spatial regions currently in the cache, and a sampling-based AGT that selectively monitors representative regions. This separation allows for precise pattern learning until the intended End-of-Generation (EoG) event when an element of the region is evicted from the cache (\Cref{ssec:jail-sampling}).

\end{itemize}

\subsection{Overview}\label{ssec:overview}
 Proba's design is given in~\Cref{fig:proba}. It inherits the overall organization of prior SMS-based prefetchers. Without loss of generality, we refer to the structure that tracks access patterns within a spatial region as the Active Generation Table (AGT), and the structure that stores historical metadata governing prefetch decisions as the Pattern History Table (PHT). Proba observes misses and prefetch tag-hits\footnote{\shepherd{As in Triangel~\cite{Triangel} and Berti~\cite{Berti}, Proba trains on misses and prefetch tag hits rather than all demand accesses. Where relevant we use the tag-based accuracy definitions from Feedback-Directed Prefetching~\cite{4147648}: A prefetched cache line has a tag associated with it in the cache. On first use, the tag is unset and the prefetch is treated as accurate. If the line is evicted while the tag remains set, the prefetch is treated as inaccurate.}} at the L2, along with evictions in order to track generations of accesses to 4KiB regions. The AGT records the access pattern and trigger feature for each active region. When the AGT finishes tracking a region and produces a new observation, \name{} estimates accuracy by comparing the observed bit pattern with the history patterns stored in the PHT. \calibration{The PHT-entry-local accuracy estimate is calibrated using global prefetch accuracy to mitigate systematic bias.} Based on this accuracy estimate, \name{} performs probability state transitions and generates a probability pair. Using this pair, \name{} probabilistically updates the history pattern based on new observations, enabling fine-grained pattern evolution without storage overhead. This update policy is applied across multiple trigger-feature-indexed PHTs, using marginal accuracy to ensure each trigger feature is only as aggressive as justified by its unique contribution. To avoid imprecise pattern learning and resource thrashing, Proba introduces a sampling-based AGT that selectively trains on representative regions, while a separate Jail Table tracks active regions currently present in the cache and ensures that prefetches are only triggered by valid access events. When an access with the same trigger feature occurs in a new region, \name{} indexes the PHTs using various trigger features and issues prefetches from the union of their stored history patterns.

\subsection{Probabilistic Update}\label{ssec:proba-update}

\begin{figure}
    \centering
    \includegraphics[width=\linewidth]{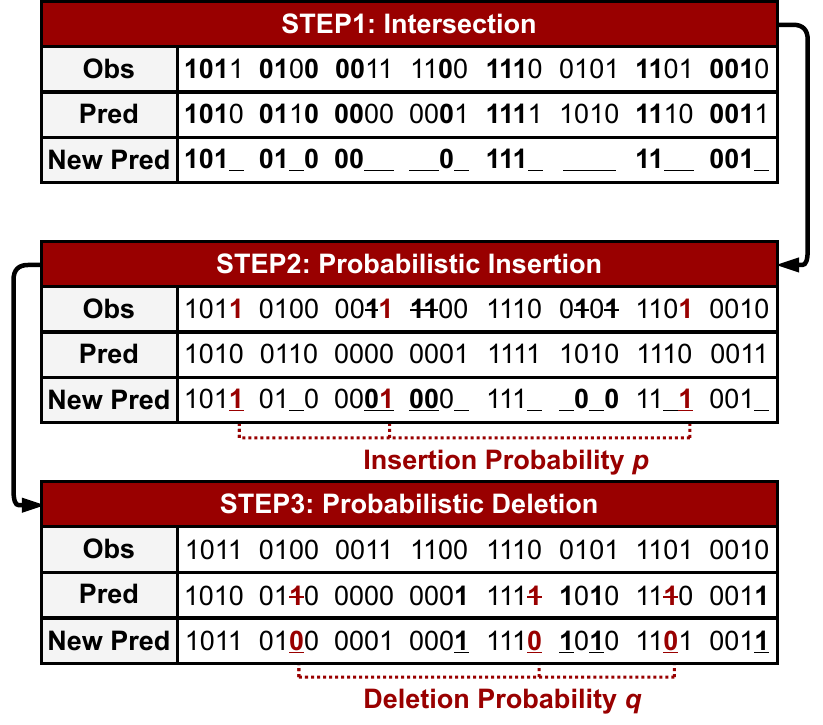}
    \vspace{-25pt}
    \caption{Probabilistic update of the Pattern History Table (PHT). Step~1 retains the intersection between the new observation (from the AGT) and the previous prediction (from the PHT). Step~2 inserts bits present in the observation but absent in the old prediction with an insertion probability $p$. Step~3 deletes bits present in the old prediction but missing in the new observation with a deletion probability $q$.}
    \label{fig:proba-update}
    \vspace{-15pt}
\end{figure}
\begin{figure*}[t]
    \centering
    \includegraphics[width=0.95\textwidth]{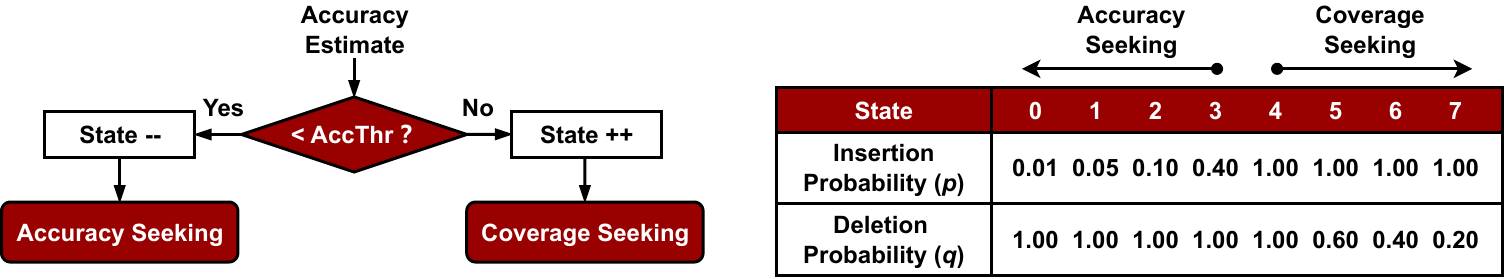}
    \vspace{-5pt}
    \caption{State transition logic for \name{} and the corresponding mapping from states to insertion and deletion probabilities.}
    \label{fig:acc-cov-values}
    \vspace{-10pt}
\end{figure*}
When the Active Generation Table (AGT) finishes tracking a spatial region, it produces an observation bit pattern $O \in \{0,1\}^L$, where \shepherd{$L$ is the number of cache lines in a spatial region (i.e., the region size divided by the cache-line size)}, and each bit indicates whether the corresponding cache line in the region was accessed. The corresponding Pattern History Table (PHT) entries, indexed by trigger feature (\Cref{ssec:marginalpht}), maintain history bit patterns $P \in \{0,1\}^L$. As illustrated in \Cref{fig:proba-update}, \name{} treats the update of a history pattern $P$ with observation $O$ as a stochastic process that interpolates between the most conservative (bitwise AND) and the most aggressive (bitwise OR) behavior. For each bit position $i \in \{1,\dots,L\}$, \name{} applies the following update rule:
\begin{equation}
P_i^{\text{ new}} =
\begin{cases}
1, \quad \text{with probability } p \quad \text{if } O_i=1 \wedge P_i=0,\\
0, \quad \text{with probability } q \quad \text{if } O_i=0 \wedge P_i=1,\\
P_i, \quad \text{otherwise.}
\end{cases}
\label{eq:proba-update}
\end{equation}
where $p \in [0.0,1.0]$ is the \emph{insertion probability} which controls how aggressively newly observed bits are incorporated into the history, and $q \in [0.0,1.0]$ is the \emph{deletion probability} which controls how aggressively bits absent in the current observation are removed. 

\ignore{Intuitively, the set intersection $P \wedge O$ yields the most accurate history (minimizing false positives), while the union $P \vee O$ maximizes coverage (minimizing false negatives). \name{} generalizes these extremes:
\begin{equation}
\begin{aligned}
(p,q) &= (0.0,1.0) &&\Rightarrow&& P^{\text{ new}} = P \wedge O,\\
(p,q) &= (1.0,0.0) &&\Rightarrow&& P^{\text{ new}} = P \vee   O
\end{aligned}
\label{eq:proba-extreme}
\end{equation}
}

Choosing a low \emph{insertion probability} $p$ and a high \emph{deletion probability} $q$ biases the update toward accuracy by retaining only persistently observed bits; conversely, a high $p$ and low $q$ biases toward coverage by quickly admitting new bits and slowly forgetting stale ones. To prevent bit patterns from tending to all 1s or 0s, \name{} avoids the extreme combinations $p=1.0$, $q=0.0$ and $p=0.0$, $q=1.0$. Over repeated observations, applying Eq. \eqref{eq:proba-update} yields a self-refreshing history pattern where fresh observations $O$ are blended into history pattern $P$ while obsolete patterns gradually decay. 
\shepherd{The probabilistic update\footnote{\shepherd{Probabilistic PHT updates occur only when an AGT observation completes, at EoG or on a premature AGT eviction (\Cref{ssec:jail-sampling}), and can therefore be buffered without delaying prefetch generation. Moreover, Proba requires only simple randomized update logic: a lightweight pseudo-random generator, such as a linear congruential generator, is sufficient. Randomized decisions are also common in practical cache-management mechanisms, including Bimodal RRIP (BRRIP)~\cite{RRIP}. We tried several random number generators during our experiments, including 
\texttt{std::rand}, \texttt{std::minstd\_rand}, \texttt{std::mt19937}, and \texttt{std::knuth\_b}, 
and changing the RNG produced no material difference in any simulated outcome in any simulated workload.}} logic is entirely off the prefetch-issue and L2-access critical paths: 
randomness is required only when committing a completed observation to the PHT.}

\subsection{Accuracy-Coverage Seeking Decision-making}\label{ssec:seeking}
\name{} dynamically adjusts its accuracy-coverage seeking behavior in response to accuracy feedback. An accuracy threshold $AccThr$, with a default value of 80\%,\ignore{\footnote{This threshold may change in response to other feedback, including DRAM bandwidth utilization, cache pollution~\cite{4147648}, and criticality~\cite{CLIP}.}} guides adaptation. On each PHT update, if the measured accuracy exceeds $AccThr$, the prefetcher transitions toward a more coverage-seeking behavior on the next update; otherwise it shifts toward a more accuracy-seeking behavior on the next update\ignore{\footnote{\justin{I am inclined to remove this footnote and keep the randomness one instead} There is one special case where the behavior does not change: when the bit pattern has no bits set other than the trigger access (i.e., $C_{pred}-1=0$). In this case no prefetches are generated and the accuracy metric is undefined, so the state remains unchanged. This prevents rapid oscillation between accuracy- and coverage-seeking, while still allowing the bit pattern to evolve: once a new bit is observed, it can be inserted because the insertion probability is nonzero in every state.}}.

Proba adopts the PHT-entry-local accuracy metric from DSPatch, which we denote as $LocalAcc$. This metric compares the new observation (\textit{O}) with the history pattern used for prediction (\textit{P}) to quantify correctness of the pattern. 
\begin{equation}
    LocalAcc = \frac{C_{acc}-1}{C_{pred}-1} 
    \label{eq:proba-accuracy-base}
\end{equation}
where $C_{acc} = |O\ \cap P|$ and $C_{pred} = |P|$, and $|X|$ is the number of elements in set $X$. The trigger access is excluded.

\name{} encodes its accuracy–coverage decision in a 3-bit field, \texttt{State}, within each PHT entry, where each probability state corresponds to a distinct pair of insertion and deletion probabilities. The state transition logic and probability map are given in \Cref{fig:acc-cov-values}. The mid-point, where $p=q=1.0$, matches the behavior of the baseline SMS, and was chosen as the best-performing unbiased probability state: lower probabilities without bias, e.g. $p=q=0.5$, performed less well. For the same reason, all coverage-seeking states use $p=1.0$ and all accuracy-seeking states use $q=1.0$. The most accuracy-seeking state has a lower insertion probability than the most coverage-seeking state's deletion probability. This design ensures that PHT indices (e.g. PCs) poorly covered by spatial memory streaming are not overfetched, while still allowing the prefetcher to quickly adapt and maintain a small chance of generating prefetches if their behavior changes in the future. \shepherd{The remaining intermediate states fill the probability space between these extremes; we tuned them on SPECspeed 2017 and found that, outside extreme settings, their exact values have little performance impact.}

\subsection{Pattern History Tables and Marginal Accuracy}\label{ssec:marginalpht}

Proba employs multiple Pattern History Tables (PHTs), each indexed by a distinct trigger feature, inspired by prior work~\cite{Gaze,SMS,Bingo,DSPatch}. The first table is indexed by trigger PC (\textbf{\texttt{PC-Only}}) using rotated indexing similar to the SMS variant~\cite{ferdman2011spatial} and DSPatch~\cite{DSPatch}. The second table is indexed by the trigger PC combined with the trigger offset (\textbf{\texttt{PC+Offset}}), like Bingo~\cite{Bingo} and the original SMS~\cite{SMS}, to capture patterns where the initial access offset is informative. The third table is indexed by the temporal correlation between the first and second offsets accessed within a region (\textbf{\texttt{Offset+Offset}}), inspired by Gaze~\cite{Gaze}. To support this, Proba uses a small First-Offset Table (FT)\footnote{The First-Offset Table (FT) stores the first accessed offset for the last 64 regions that reached the SoG event, regardless of whether they were sampled by the AGT or placed in the Jail on the first access to the region (\Cref{ssec:jail-sampling}). If a second offset is accessed within that region before the region reaches EoG and before it is evicted from the FT due to capacity pressure, we trigger \texttt{Offset+Offset} prefetching. For sampled regions that update the PHT, the AGT also specifies the second offset accessed rather than just the first.} to record the first offset observed for a region. When a distinct second offset is later accessed in the same region, the pair of offsets is used to trigger \texttt{Offset+Offset} prefetching. At prefetch time, all PHTs are queried using the corresponding trigger features. Similar to PMP~\cite{PMP}, Proba issues prefetches from the union of their predictions.

While combining multiple trigger features in this way improves coverage, it can also reduce accuracy because some trigger features may introduce incorrect prefetches. Proba addresses this through an accuracy-coverage trade-off based on \emph{marginal accuracy}. During PHT updates, trigger features are ranked by prediction accuracy. The most accurate trigger feature is updated using its absolute accuracy, while less accurate trigger features are evaluated only on the additional coverage they provide beyond more accurate trigger features. Specifically, for two trigger features $A$ and $B$, where $B$ is less accurate, we compute the marginal accurate-prefetch and marginal prediction sets as $C_{macc}[B] = C_{acc}[B] \wedge (\neg C_{pred}[A])$ and $C_{mpred}[B] = C_{pred}[B] \wedge (\neg C_{pred}[A])$, and use their ratio to determine whether to increase or decrease $B$'s aggressiveness. If the accuracy of prefetches issued by $B$ but not by $A$, considering only accesses not already covered by $A$, is above the accuracy target (80\% by default), we increase $B$'s aggressiveness; otherwise, we decrease it. This formulation extends to additional trigger features by excluding predictions already covered by more accurate trigger features. As a result, each trigger feature is rewarded only for the unique, correct prefetches it contributes, allowing Proba to improve coverage while maintaining high accuracy.

\ignore{
\subsection{Pattern History Table (PHT) Tagging}\label{ssec:pht-tagging}
Unlike DSPatch, Proba employs PHT tagging because experimentally it offers consistent improvements by reducing destructive aliasing. This is particularly important for probabilistic accuracy-seeking updates: when two unrelated spatial patterns alias to the same PHT entry, their conflicting updates drive the stored bit pattern toward low coverage, effectively collapsing the prediction.

Proba's PHT is a 16-way set-associative structure with 256 total entries and a Least Recently Used Replacement Policy (LRURP). Each entry stores history metadata together with a 12-bit tag. On each trigger access to a spatial region, the trigger PC is hashed using a folded XOR to generate a 16-bit PC signature, which is stored in the corresponding AGT entry. When updating the PHT, the lower 4 bits of this signature select the PHT set, while the upper 12 bits form the tag that is compared against the tags stored in that set. If the tag matches, the corresponding PHT entry is updated; otherwise, a victim PHT entry is selected for replacement according to the replacement policy. This tagging mechanism preserves locality for related PCs while isolating unrelated ones, reducing aliasing and allowing probabilistic updates to converge to meaningful patterns.
}

\subsection{Jail Table and Region Sampling}\label{ssec:jail-sampling}

\begin{figure}[t]
    \centering
    \includegraphics[width=\columnwidth]{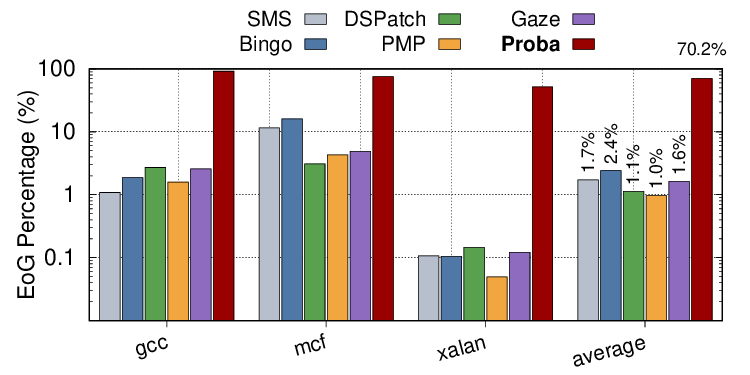}
    \vspace{-19pt}
    \caption{Percentage of Active Generation Table (AGT) entries that reach End-of-Generation (EoG) across SMS-based prefetchers on SPECspeed 2017.}
    \label{fig:eog}
    \vspace{-10pt}
\end{figure}

Experimentally, we discovered that prior SMS-based prefetchers all suffer from imprecise Start-of-Generation (SoG) and End-of-Generation (EoG) tracking due to the finite capacity of the AGT. When deployed at a large cache level (e.g., L2), only a small fraction of entries reach EoG, where a block from the corresponding region is evicted from the cache (\Cref{fig:eog}). Most entries are instead evicted prematurely from the AGT, regardless of whether the prefetcher uses a filter table~\cite{SMS} to filter single-access regions. This causes the AGT to misinterpret later accesses to the same region as new trigger accesses. As a result, what should be a single generation is fragmented into several disjoint segments, and no individual update reflects the full sequence of accesses from the true SoG to the EoG event. This fragmentation reduces coverage in the PHT and corrupts accuracy estimation because useful prefetches that would have been accurate are misclassified, and it also leads to prefetches being issued at incorrect trigger accesses, causing the prefetcher to have misplaced aggression.

\begin{figure}
    \centering
    \includegraphics[width=0.76\columnwidth]{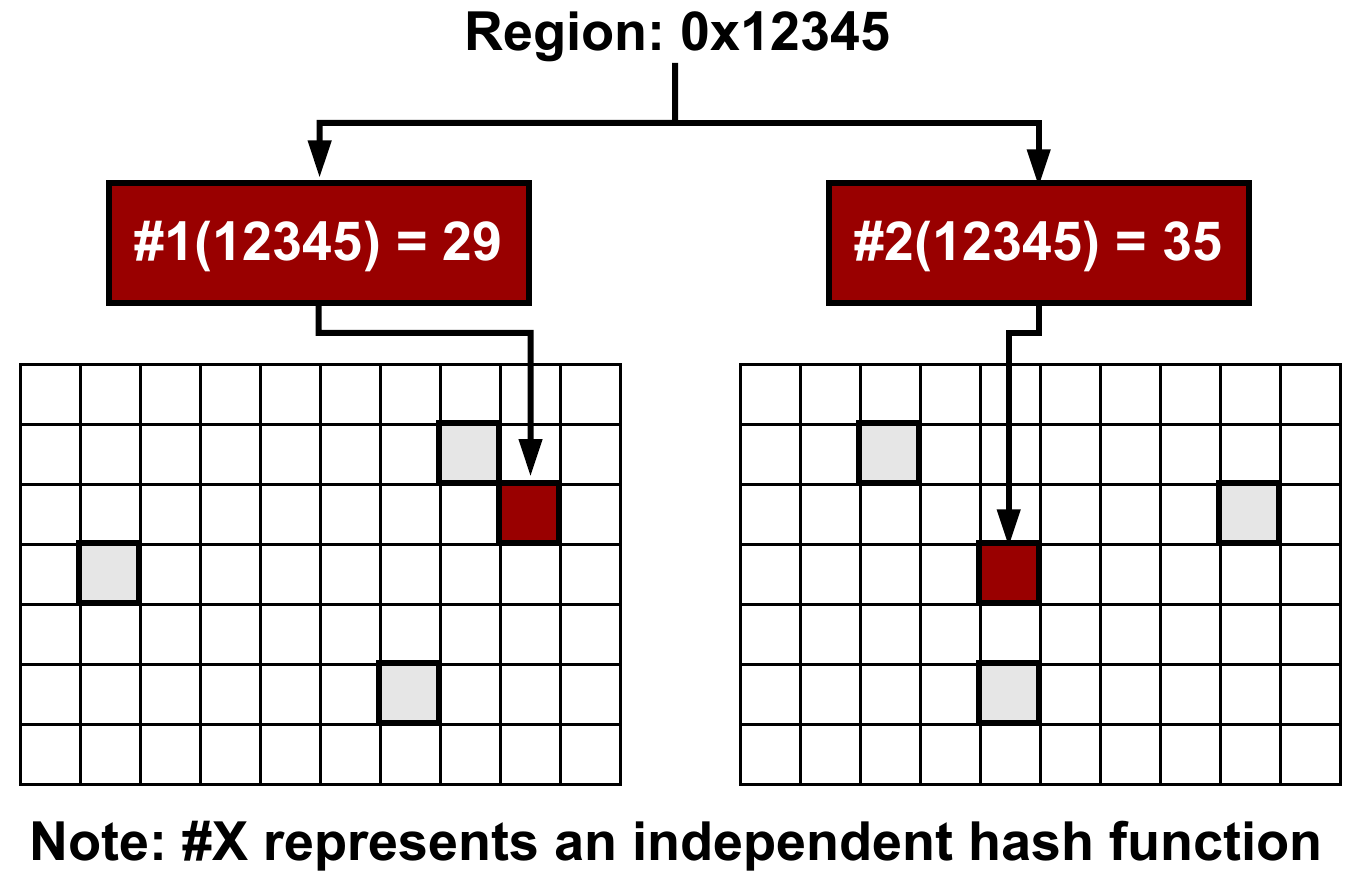}
    \vspace{-5pt}
    \caption{The Jail Table (JT) consists of two bitmaps indexed by different hash functions. When a region $r$ is inserted, bits $\#1(r)$ and $\#2(r)$ are set in bitmap~1 and bitmap~2 respectively; both bits are cleared on EoG. A region is considered live in the cache only if both bits are set, in which case no prefetches are generated and no AGT entry is allocated.}
    \label{fig:jail-table}
    \vspace{-10pt}
\end{figure}

\begin{figure}
    \centering
    \includegraphics[width=\columnwidth]{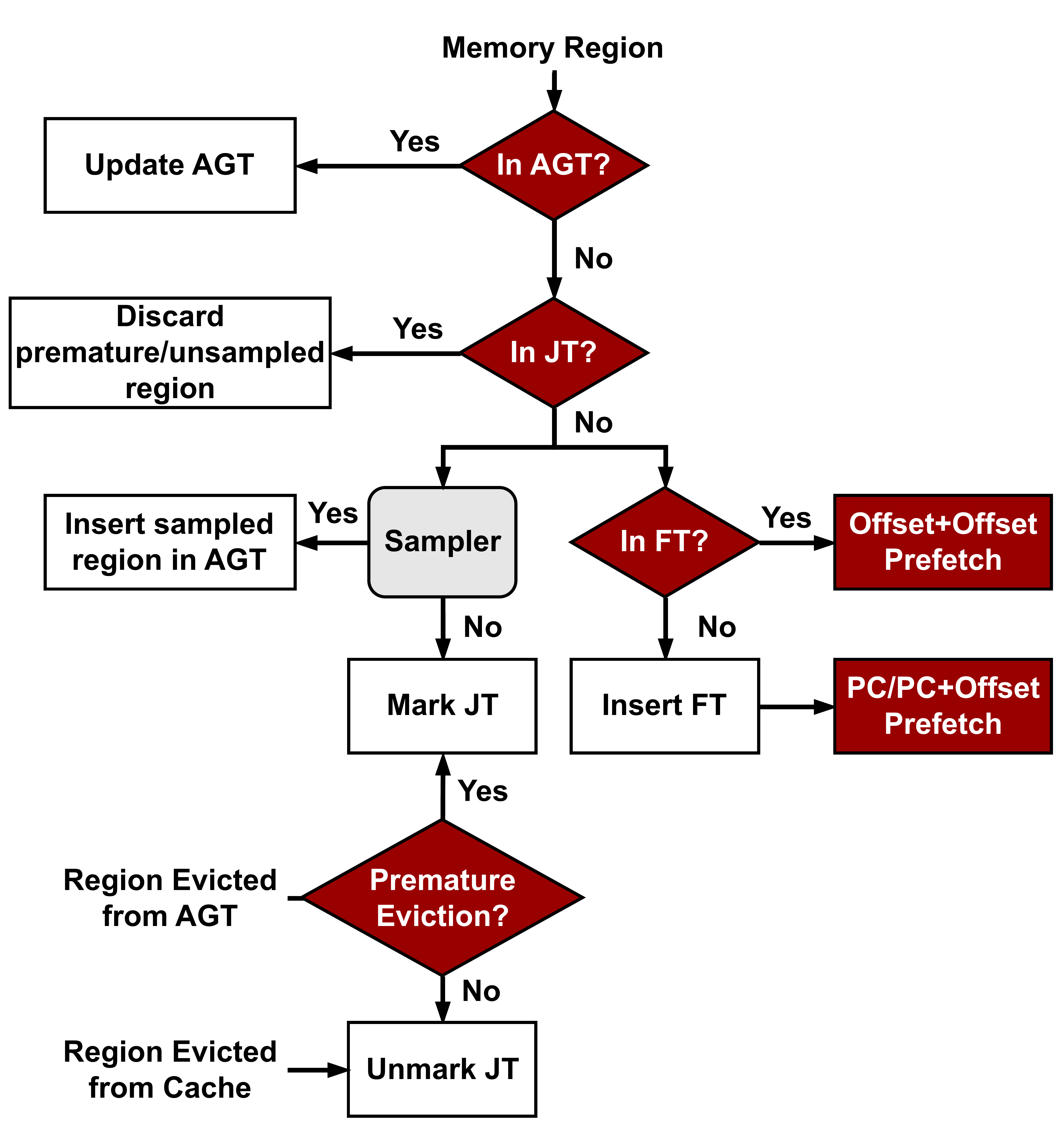}
    \vspace{-22pt}
    \caption{Sampling logic in Active Generation Table (AGT) and its interaction with the Jail Table (JT) and First-Offset Table (FT). If a region is already present in the AGT, all subsequent accesses update its bit pattern. If the region is not in the AGT but is marked in the JT (both bitmap bits set), the observation is ignored and no AGT entry or prefetch is generated until a later EoG event clears the JT entry. Otherwise, the region is potentially sampled and inserted into the AGT; regions not sampled are marked in the JT to prevent further AGT insertions. Any premature AGT eviction (before the region reaches its EoG) also marks its JT entry.}
    \label{fig:jail-sampling}
    \vspace{-10pt}
\end{figure}

To address this, we divide the functionality of the AGT into two structures: one that tracks the interval between SoG and EoG, and another that records the access pattern observed within that interval. The interval-tracking requires only minimal storage, while the pattern-tracking does not need to capture every region because access patterns are ultimately aggregated in the PHT. This allows us to reduce the required capacity through statistical sampling.

For SoG-to-EoG tracking, we introduce a new structure called the Jail Table (JT). The JT (\Cref{fig:jail-table}) stores less information than the AGT\footnote{\shepherd{The percentage of entries that reach EoG rises from 2.7\% with a 64-entry AGT but no JT to 6.1\% with 128 entries and 53.5\% with 512 entries. Proba reaches 70.2\% using a 64-entry AGT with the JT. Increasing AGT capacity is less storage-efficient because it replicates pattern-tracking state solely to preserve SoG-to-EoG intervals.}}, allowing it to track more regions with a small storage budget. The JT achieves this by only tracking whether a region is currently active (between SoG and EoG) in the cache, rather than recording its fine-grained access pattern. Upon a premature AGT eviction, the corresponding region is marked in the JT to ensure that subsequent accesses to this region are not reinserted into the AGT. The region is unmarked in the JT upon an EoG event (i.e., when any cache line of the region is evicted from the cache). When a region is marked in the JT, no prefetches are generated from it and no AGT entry is allocated to record its access pattern\ignore{\footnote{We found this worked well in all scenarios except for the most coverage-seeking state, where partial updates from non-EoG evictions increased coverage, since elements were inserted but only deleted at very low probabilities, and accuracy was already high. To handle this case, we use a simple cutoff: if global accuracy is above 90\%, we use DSPatch's policy of updating on every eviction, ignoring the Jail Table and inserting every SoG event into the AGT.}}.

To track this, we use a pair of bitmaps inspired by a Bloom filter~\cite{bloom}\footnote{The Jail Table differs from a Bloom filter, which uses a single bit array indexed by multiple hashes and does not support deletion without clearing or augmenting the structure, and from Bloom-like structures such as Cuckoo filters~\cite{10.1145/2674005.2674994}, which organize state differently. The Jail Table supports efficient deletion and deliberately permits both false positives and false negatives. Because deleting an entry removes the corresponding element from each array, it may also remove an element associated with another still-live region, thereby introducing a false negative. False negatives merely cause region splitting, whereas false positives reduce prefetch coverage.}. Two bitmaps are indexed by separate hash functions, $\#1$ and $\#2$. When a region number $r$ is inserted (at SoG), bit $\#1(r)$ is set in bitmap~1 and $\#2(r)$ is set in bitmap~2. Deletion (at EoG) clears these two bits. To test whether a region is within the SoG-EoG interval, we check whether bitmap~1 contains bit $\#1(r)$ and bitmap~2 contains bit $\#2(r)$.

To increase the number of regions that reach EoG and thereby generate frequent PHT updates, the AGT switches to \textit{sampling} under pressure (when a set reaches 50\% occupancy): new regions are inserted into the AGT with a sampling rate of 5\% (\Cref{fig:jail-sampling}). To prevent accesses to unsampled regions from entering the AGT, we reuse the JT: such regions are marked in the JT at first access, rather than only upon premature AGT evictions. If a region is already marked in the JT, it is not allocated to the AGT again until its EoG event is observed. Unlike SMS and DSPatch, Proba employs a Random Replacement Policy (RandomRP) in the AGT to support unbiased sampling. This sampling significantly reduces AGT traffic while providing a sufficiently broad set of observations for high-quality PHT updates.

The PHTs are updated when either an EoG event occurs in the cache or the AGT prematurely evicts an entry, but due to sampling, these premature evictions are now much rarer and comparatively more AGT entries reach the intended EoG point (\Cref{fig:eog}). 

\begin{table}[t]
\renewcommand{\arraystretch}{1.5}
\centering
\caption{Detailed storage requirements of Proba}
\label{tab:storage_requirements}
\vspace{-5pt}
{
\begin{tabular}{|p{0.5in}<{\centering}|p{1.7in}|p{0.7in}<{\centering}|}
\hline
\textbf{Structure} &
\centering\textbf{Description} &
\textbf{Storage} \\ \hline\hline

FT &
\makecell[l]{
\specialrule{0em}{0.5pt}{0.5pt}
$8$-way; $64$-entry, LRURP \\
$\bullet$ Region Tag ($36$b) \\
$\bullet$ Trigger Offset ($6$b) \\
$\bullet$ LRU State ($3$b) \\
\specialrule{0em}{0.5pt}{0.5pt}
} &
$360$B \\ \hline

AGT &
\makecell[l]{
\specialrule{0em}{0.5pt}{0.5pt}
$8$-way; $64$-entry, RandomRP\\
$\bullet$ Region Tag ($36$b) \\
$\bullet$ Trigger PC ($16$b) \\
$\bullet$ Trigger Offset ($6$b) \\
$\bullet$ Second Offset ($6$b) \\
$\bullet$ Observation Bit Pattern ($64$b) \\
\specialrule{0em}{0.5pt}{0.5pt}
} &
$1024$B \\ \hline

JT &
\makecell[l]{
\specialrule{0em}{0.5pt}{0.5pt}
Bloom-like filter, dual hash\\
$\bullet$ Array ($2 \times 4096$b) \\
\specialrule{0em}{0.5pt}{0.5pt}
} &
$1024$B \\ \hline

PHT &
\makecell[l]{
\specialrule{0em}{0.5pt}{0.5pt}
16-way, $256$-entry, BRRIPRP\\
$\bullet$ Tag ($12$b) \\
$\bullet$ Proba Bit Pattern ($64$b) \\
$\bullet$ State ($3$b) \\
$\bullet$ BRRIP State ($3$b) \\
\specialrule{0em}{0.5pt}{0.5pt}
} &
$3 \times 2624$B \\ %
\hline\hline

\multicolumn{2}{|c|}{\textbf{Total}} & \textbf{$\mathbf{10.04}$KiB} \\ \hline
\end{tabular}
}
\end{table}

\subsection{Hardware Overhead}\label{ssec:hardware-overhead}
 Overall, Proba requires 10.04 KiB of storage (\Cref{tab:storage_requirements}). Proba tracks 4 KiB regions, which correspond to the size of a typical physical page. With 64-byte cache lines, each region contains $4096/64=64$ cache-line offsets, so each spatial pattern is represented by a 64-bit pattern. Each region entry also stores a 36-bit region tag, a 6-bit trigger offset, and, where required, a 6-bit second offset. The First-Offset Table (FT) stores only the region tag and trigger offset, since it is used only to trigger the \texttt{Offset+Offset} trigger feature. The Active Generation Table (AGT) stores both the trigger and second offsets so that the \texttt{Offset+Offset} trigger feature can be trained when an entry is evicted. Both FT and AGT are 64-entry, 8-way set-associative structures; FT uses LRU replacement, while AGT uses Random Replacement Policy (RandomRP). Pattern History Tables (PHTs) support three trigger features: \texttt{PC-Only}, \texttt{PC+Offset}, and \texttt{Offset+Offset}. Each PHT is a 256-entry, 16-way set-associative structure using BRRIP replacement. The Jail Table (JT) uses two hash functions and two 4096-bit bitmaps.

\section{Experimental Setup}

\begin{figure*}[t]
    \centering
    \subfloat{
\resizebox{\linewidth}{!}{%
    \includegraphics{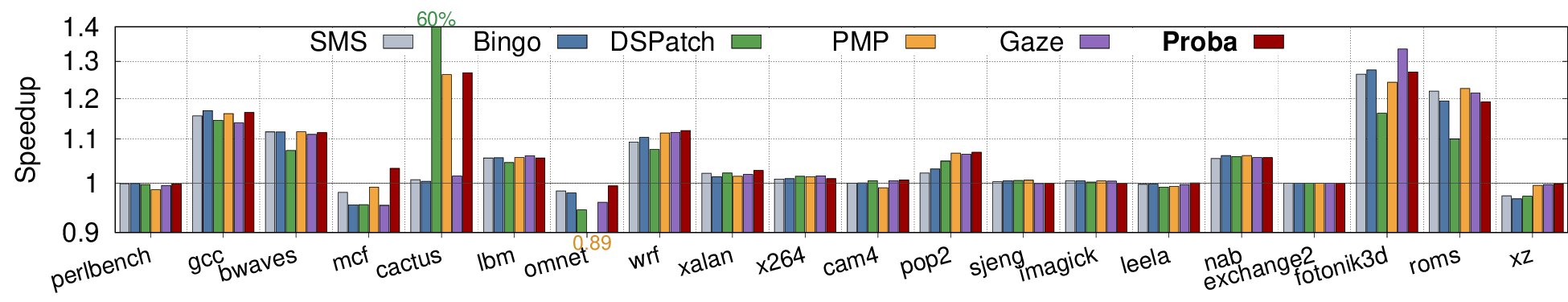}
    \includegraphics{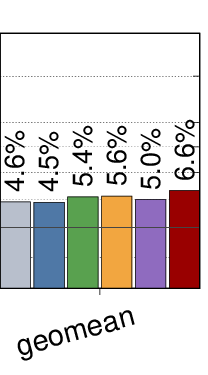}
}
            \vspace{-5pt}
    \label{fig:speedup17}
   }
   
    \subfloat{
\resizebox{\linewidth}{!}{%
    \includegraphics{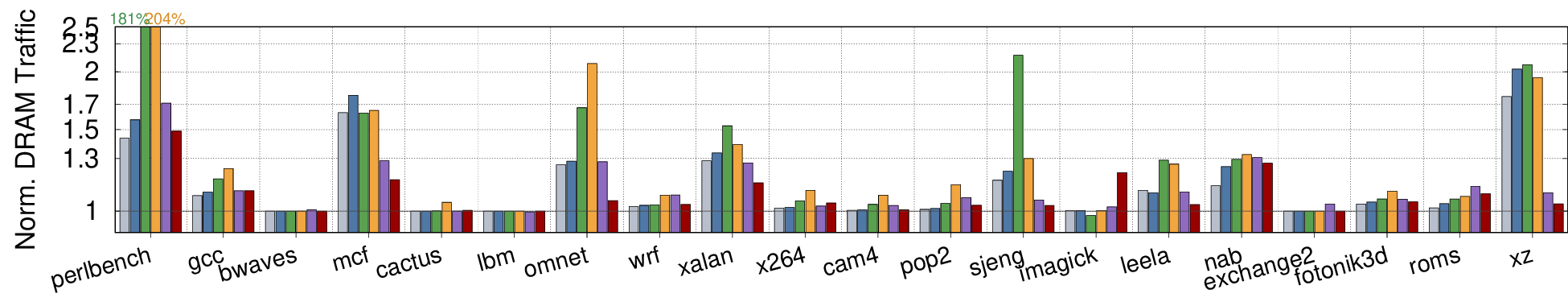}
    \includegraphics{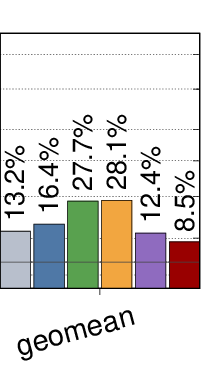}
}
    \label{fig:dram17}
   }  
   \vspace{-10pt}

   \caption{Single-core performance and normalized DRAM traffic of the evaluated SMS-based prefetchers on SPECspeed 2017.}
    \label{fig:speedupdram}
            \vspace{-10pt}
\end{figure*}

\ignore{
\begin{table}[t]\renewcommand{\arraystretch}{1.5}
    \centering
    \caption{Configuration of the gem5 simulator}
    \label{tab:simulator_configuration}
    {
        \begin{tabularx}{3.3in}{m{0.35in} m{2.5in}}
            \hline
            \textbf{Core} & 5-Wide, out-of-order, 2GHz \\
            \hline
            \textbf{Pipeline} & 288-Entry ROB, 120-ent IQ, 85-ent LQ, 90-ent SQ, 150 Int/256 FP regs, 4 Int ALUs, 2 Mul/Div, 4 FP/SIMD, 2 R/W \\
            \hline
            \textbf{Branch Pred.}  & 64KiB MPP-TAGE \\
            \hline
            \textbf{L1I}  & 64KiB, 4-way, 2-cycle hit latency, 16 MSHRs \\
            \hline
            \textbf{L1D}  & 64KiB, 4-way, 4-cycle hit latency, 16 MSHRs, deg-8 stride PF \\
            \hline
            \textbf{L2C}  & 512KiB, 8-way, 9-cycle hit latency, 32 MSHRs \\
            \hline
            \textbf{LLC}  & 2MiB/core, 16-way, 20-cycle hit latency, 36 MSHRs \\
            \hline
            \textbf{DRAM} & LPDDR5\_5500\_1x16\_BG\_BL32 \\
            \hline
            \textbf{OS} & Ubuntu 22.04 \\
            \hline
        \end{tabularx}
    }
\end{table}
}

\begin{table}[t]\renewcommand{\arraystretch}{1.2} 
\caption{Simulation parameters for baseline system}
\vspace{-5pt}
\label{tab:simulator_configuration}
\begin{center}
    \begin{tabular}{|p{1.2cm}|p{6.5cm}|}
    \hline  
\textbf{Core} & Out-of-order, {TAGE-SC-L} Branch Pred., 4 GHz, 8-issue width,
8-retire width, 576-entry ROB \\ \hline
\textbf{TLBs} & L1 iTLB \& dTLB:  64 entries, 4-way, 1 cycle 

STLB: 1536 entries, 12-way, 8 cycles \\\hline
\textbf{L1I} & 64 KiB, 8-way, 4 cycles \\\hline
\textbf{L1D} & 48 KiB, 12-way, 5 cycles, LRURP \\ \hline
\textbf{L2C} & {2 MiB}, 16-way, 10 cycles, LRURP \\\hline
\textbf{LLC} & {3 MiB / core}, 12-way, 35 cycles, DRRIPRP \\\hline
\textbf{MSHRs} & 8/16/32/64 in L1I/L1D/L2/LLC \\\hline
\textbf{PQs} & 32/8/16/32 in L1I/L1D/L2/LLC \\\hline
\textbf{DRAM} &
128GB, 2400 MTPS (4800 multicore), 8 banks
\\ \hline
\end{tabular}
\end{center}
\vspace{-10pt}
\end{table}

\begin{table}[t]
\renewcommand{\arraystretch}{1.5}
\centering
\caption{Parameters of each comparison prefetcher}
\label{tab:all_storage_requirements}
\vspace{-5pt}
{
\begin{tabular}{|p{0.6in}<{\centering}|p{1.82in}|p{0.45in}<{\centering}|}
\hline
\textbf{Name} &
\centering\textbf{Parameters} &
\textbf{Storage} \\ \hline\hline

SMS~\cite{SMS} &
\makecell[l]{
\specialrule{0em}{0.5pt}{0.5pt}
4KiB region, 64-entry AGT, 1024-\\
entry PHT,
\textbf{Index: \texttt{PC+Offset}} \\
\specialrule{0em}{0.5pt}{0.5pt}
} &
$11.95$KiB
\\ \hline

Bingo~\cite{Bingo} &
\makecell[l]{
\specialrule{0em}{0.5pt}{0.5pt}
4KiB region, 64-entry AGT,\\
1024-entry PHT, \\
\textbf{Index: \texttt{PC+Offset/PC+Addr}} \\
\specialrule{0em}{0.5pt}{0.5pt}
} &
$13.20$KiB
\\ \hline

DSPatch~\cite{DSPatch} &
\makecell[l]{
\specialrule{0em}{0.5pt}{0.5pt}
4KiB region, 64-entry Page Buf.,\\
1024-entry SPT,
\textbf{Index: \texttt{PC-Only}} \\
\specialrule{0em}{0.5pt}{0.5pt}
} &
$9.68$KiB
\\ \hline

PMP~\cite{PMP} &
\makecell[l]{
\specialrule{0em}{0.5pt}{0.5pt}
4KiB region, 64-entry AGT,\\
64-entry OPT, 256-entry PPT, \\
\textbf{Index: \texttt{Offset/PC-Only}} \\
\specialrule{0em}{0.5pt}{0.5pt}
} &
$9.02$KiB
\\ \hline

Gaze~\cite{Gaze} &
\makecell[l]{
\specialrule{0em}{0.5pt}{0.5pt}
4KiB region, 64-entry AGT,\\
1024-entry PHT, 8-entry DPCT, \\
\textbf{Index: \texttt{Offset+Offset}} \\
\specialrule{0em}{0.5pt}{0.5pt}
} &
$10.81$KiB
\\ \hline

\end{tabular}
\vspace{-10pt}
}
\end{table}

We evaluate \name{} with the trace-based ChampSim simulator~\cite{gober2022championshipsimulatorarchitecturalsimulation}, using the configuration in \Cref{tab:simulator_configuration}. We compare \name{} with several state-of-the-art SMS-based prefetchers, with detailed configurations reported in \Cref{tab:all_storage_requirements}. For fairness, all evaluated prefetchers are placed at the L2 cache, track 4 KiB spatial regions, and are configured to use a comparable storage budget. All configurations, including the baseline, include a degree-3 IP-stride prefetcher~\cite{381947} at the L1 cache. Prefetcher implementations are taken from the original authors~\cite{Gaze,PMP,Bingo,DSPatch,SMS}.

\shepherd{The single-core configuration uses one DRAM channel at 2400 MTPS. Multicore experiments use a data rate of 4800 MTPS and two DRAM channels for both the 2-core and 4-core configurations; the 4-core configuration additionally uses two ranks per channel. This configuration is inspired by the one used in Gaze ~\cite{Gaze}.}

We evaluate on simpoint traces from four benchmark suites: SPECspeed 2017 with 95 traces, SPEC CPU2006 with 93 traces, Ligra with 76 traces, and CloudSuite with 120 traces. \shepherd{For each trace, we use a 200-million-instruction warmup followed by a 200-million-instruction simulation window.} Results are aggregated across simpoints of each workload, with weights where available.

\section{Evaluation}

\subsection{Single Core Analysis on SPECspeed 2017}

In the breakdown of SPECspeed 2017 (\Cref{fig:speedupdram}), we show that Proba delivers a 6.6\% geomean speedup while incurring only 8.5\% additional DRAM traffic. PMP~\cite{PMP} is the next-best design in terms of performance, at 5.6\% speedup, but does so with a substantially higher 28.1\% DRAM traffic overhead. This is despite PMP using large 5-bit saturating counters to track patterns across historical updates, whereas Proba uses just a single probabilistic bit to do the same. Indeed, PMP's high traffic causes slowdowns in \texttt{omnet} and \texttt{cam4}, where Proba instead behaves in an accuracy-seeking manner. PMP also slightly slows \texttt{mcf} by 0.8\%, whereas Proba improves its performance by 3.3\%, while generating substantially less traffic. In our setup, Gaze~\cite{Gaze} is less performant than PMP, achieving a 5.0\% speedup. Although its traffic is considerably more modest than PMP's, at 12.4\% above baseline, it is still less frugal than Proba. We attribute this to Gaze's reliance on a single \texttt{Offset+Offset} trigger feature, which reduces coverage in workloads such as \texttt{gcc} and \texttt{mcf}. While Gaze includes an integrated stride prefetcher, its benefit is limited due to overlap with the L1 stride prefetcher in the baseline. In some cases, Gaze outperforms Proba by being more aggressive, but more often the resulting increase in DRAM traffic leads to lower overall performance, even in this single-threaded setting with ample bandwidth. DSPatch, Bingo, and SMS also underperform Proba in geomean speedup while generating substantially more DRAM traffic. A particularly strong per-workload exception is \texttt{cactuBSSN}, where DSPatch's pattern-compression mechanism (\Cref{sec:dspatch_background}) happens to increase coverage with little accuracy loss. This benefit is specific to compression rather than DSPatch's underlying PHT update policy, and disappears when DSPatch is evaluated without compression. As previously shown in \Cref{fig:summary-means}, the performance gap grows when DRAM bandwidth is reduced relative to default. As we shall see, this is also reflected when effective bandwidth is reduced by contention in multicore workloads.

\shepherd{Beyond these performance and traffic trends, we can also consider prefetcher behavior over phase transitions. For example, under Proba, the IPC over 10-million-instruction intervals for the \texttt{wrf\_s-6673B} trace drops sharply from 3.62 in one interval to 0.90 in the next, while the \texttt{pop2-17B} trace drops from 2.25 in one interval to 1.57 in the next. Despite these abrupt changes in workload behavior, Proba achieves the highest speedup among the evaluated prefetchers on both workloads, suggesting that its probabilistic adaptation remains effective across sharp phase transitions.}

\begin{figure}[t]
    \centering

    \includegraphics[width=\linewidth]{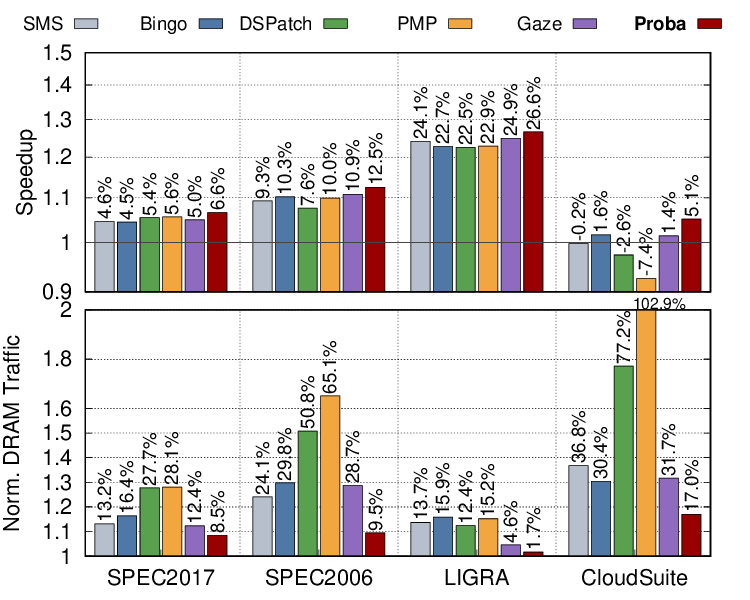}
    \vspace{-17pt}
    \caption{Single-core speedup and normalized DRAM traffic of prefetchers on various benchmark suites (geomean).}
    \label{fig:benches-all}
    \vspace{-10pt}
\end{figure}

\subsection{Alternative Benchmark Suites}
In other suites, Proba's DRAM traffic reduction translates into larger performance gains. \Cref{fig:benches-all} shows that, on SPEC CPU2006, Proba outperforms the closest competitor, Gaze, by 1.7\% while incurring roughly one third of its additional DRAM traffic, and outperforms PMP by 2.6\%. Proba also achieves large gains on workloads such as \texttt{mcf}, where it improves performance by 29.7\%, compared with 26.0\% for DSPatch and substantially smaller speedups for the other prefetchers. This suggests that Proba's multiple trigger features help distinguish useful spatial patterns that other designs fail to capture accurately. Ligra also benefits substantially from Proba, largely due to its very low traffic overhead of only 1.7\%. The advantage is even more pronounced on CloudSuite: Proba achieves a 5.1\% speedup above baseline, outperforming every competitor across all workloads. In these data-center workloads, for which SMS was originally designed~\cite{SMS}, Proba is not only more accurate but also achieves higher coverage than other SMS-based prefetchers.

\begin{figure}[t]
    \centering

    \includegraphics[width=\linewidth]{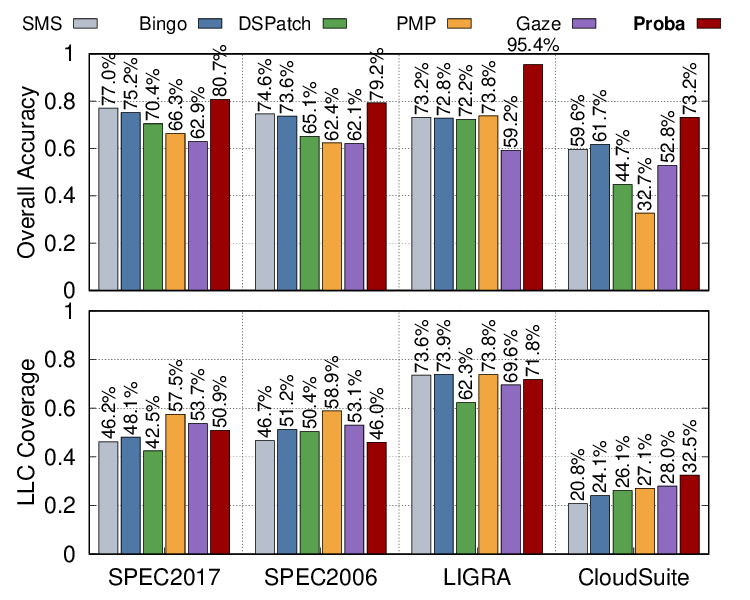}
    \vspace{-18pt}
    \caption{Overall Accuracy and LLC Coverage of prefetchers on various benchmark suites (average).}
    \vspace{-15pt}
    \label{fig:benches-acccov}
\end{figure}

Indeed, \Cref{fig:benches-acccov} shows that Proba consistently achieves the highest overall prefetch accuracy\footnote{Overall prefetch accuracy is calculated as (L2 useful prefetches + LLC useful prefetches) / (L2 useful prefetches + LLC useful prefetches + L2 useless prefetches + LLC useless prefetches). This accounts for prefetchers such as PMP and Gaze, which can prefetch into both the L2 and LLC.}, as expected from its low DRAM traffic. LLC coverage\footnote{LLC coverage is calculated as 1 - (LLC demand misses with prefetching / LLC demand misses without prefetching).} is sometimes lower than that of competing designs, but this is intentional: when a workload is poorly suited to SMS and accurate predictions are difficult, Proba reduces coverage to avoid harmful DRAM traffic, preserving performance rather than pursuing aggressive but low-value prefetches.

\begin{figure}
    \centering

    \includegraphics[width=\linewidth]{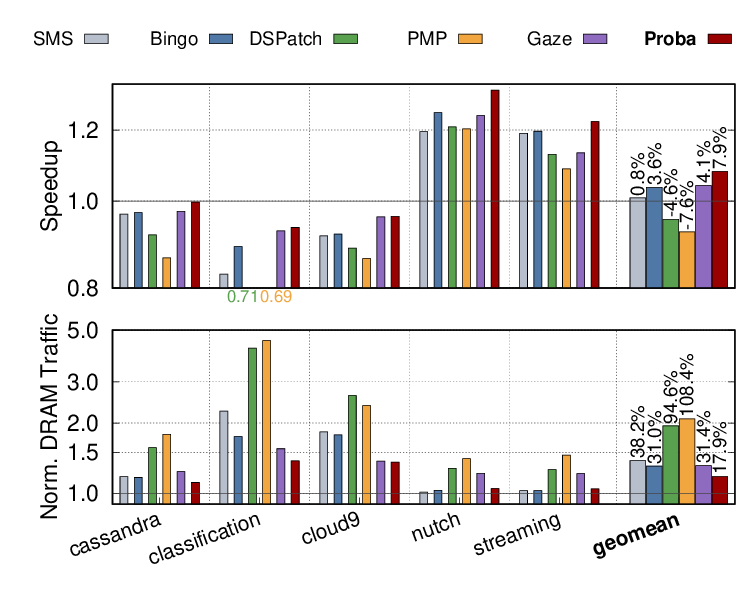}
    \vspace{-25pt}
    \caption{Speedup and DRAM traffic for 4-core CloudSuite.}
    \vspace{-10pt}
    \label{fig:4core-cloudsuite}
\end{figure}

\subsection{Multicore Performance Analysis}
Proba's reduced traffic become even more prominent in multicore settings, where the shared DRAM becomes a critical bottleneck. \Cref{fig:4core-cloudsuite} shows a breakdown of the workloads in CloudSuite, where Proba achieves a geomean speedup of 7.9\% for 4-core execution, almost double that of the nearest contender. DSPatch and PMP suffer significant slowdowns in \texttt{classification} and \texttt{cloud9}, as their higher traffic becomes harmful under contention. While Proba also sees slight slowdowns in these workloads, they are outweighed by substantial gains in \texttt{nutch} and \texttt{streaming}.

\begin{figure}
    \centering
    \includegraphics[width=\linewidth]{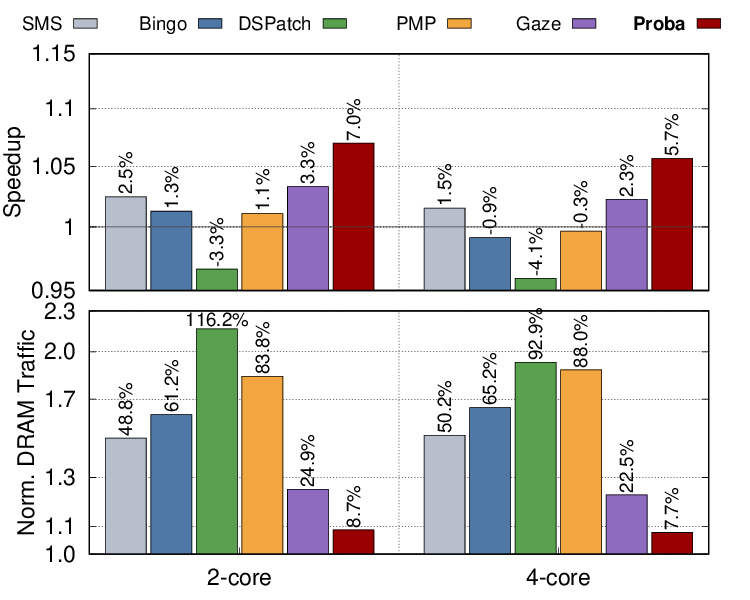}
    \vspace{-20pt}    
    \caption{Geomean speedup and normalized DRAM traffic for 2-core and 4-core homogeneous mixes of memory-intensive SPECspeed 2017 and SPEC CPU2006 workloads.}
    \label{fig:multispecmix}
    \vspace{-15pt}
\end{figure}

\Cref{fig:multispecmix} shows homogeneous mixes of the same memory-intensive SPEC workloads (\texttt{mcf}, \texttt{gcc}, \texttt{lbm}, \texttt{omnet} and \texttt{xalan}, all from both 2006 and 2017) running on 2 and 4 cores. While Proba’s advantage over PMP is moderate in the single-core setting (\Cref{fig:speedupdram}), it becomes much larger under multicore execution (5.7\% vs.\ -0.3\% for 4-core, and 7.0\% vs.\ 1.1\% for 2-core). This is because the additional DRAM traffic from most prefetchers becomes increasingly detrimental as cores scale. Among all designs, only Gaze and Proba maintain reasonable traffic levels; however, Gaze still underperforms (2.3\% for 4-core and 3.3\% for 2-core) due to lower coverage in performance-critical workloads.

\subsection{Ablation Study}
\label{ssec:ablation}
\begin{figure}
    \centering
    \includegraphics[width=\columnwidth]{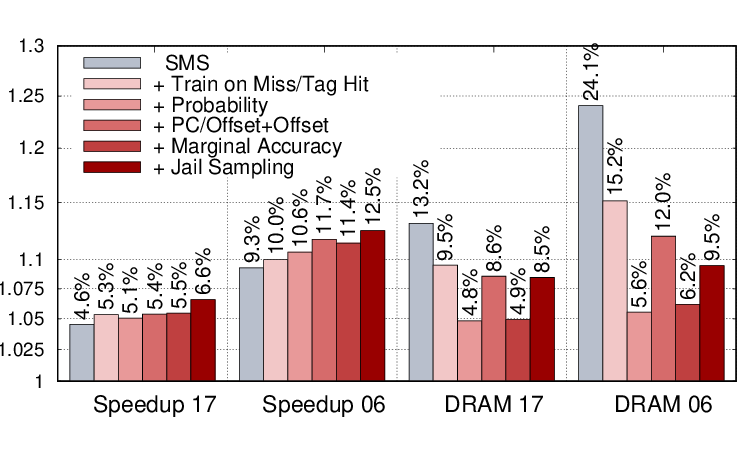}
    \vspace{-25pt}
    \caption{Ablation study. Speedup and DRAM traffic shown for SPECspeed 2017 and SPEC CPU2006 (geomean).}
    \label{fig:ablation}
    \vspace{-10pt}
\end{figure}

\Cref{fig:ablation} shows how components of Proba contribute to the performance and DRAM traffic, starting from an SMS baseline. Training on L2 misses and prefetch tag hits (\Cref{ssec:overview}) improves DRAM efficiency and is essential for Proba's accuracy targeting, as it prevents redundant cache hits from being misclassified as accurate prefetches. The dynamic probabilistic update mechanism (\Cref{ssec:proba-update} and \Cref{ssec:seeking}), which is the core of Proba’s design, is crucial for reducing DRAM traffic while maintaining competitive speedup. Introducing multi-feature Pattern History Tables with \texttt{PC-Only} and \texttt{Offset+Offset} trigger features improves coverage by capturing a wider range of spatial correlations. Adding marginal accuracy decision-making (\Cref{ssec:marginalpht}) further improves the balance between speedup and traffic by filtering predictions based on their expected benefit. Finally, the Jail Table and region sampling (\Cref{ssec:jail-sampling}) provide more complete observations, improve Proba’s accuracy estimates, and help bring prefetch aggressiveness closer to the desired level, boosting speedup while moderately increasing DRAM traffic.

\begin{figure}
    \centering
    \includegraphics[width=\columnwidth]{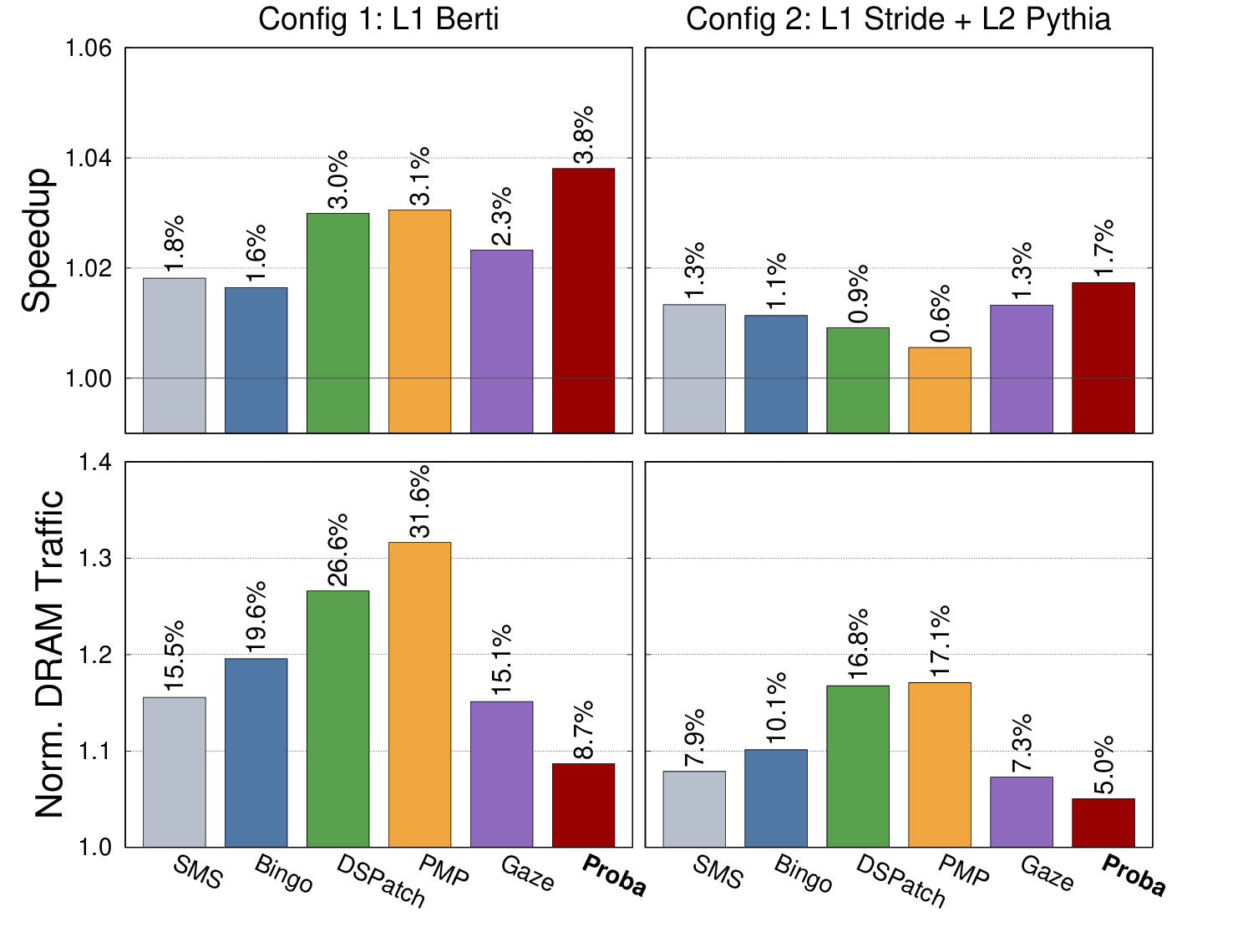}
    \vspace{-25pt}
    \caption{\shepherd{Geomean speedup and normalized DRAM traffic on SPECspeed 2017 for two multi-prefetcher configurations: Config 1 uses Berti at L1
    with an SMS-based prefetcher at L2, while Config 2 uses a
    degree-3 stride prefetcher at L1 and Pythia together with an
    SMS-based prefetcher at L2.}}
    \label{fig:hybrid-configs-ipcdram}
    \vspace{-10pt}
\end{figure}

\subsection{\shepherd{Integration with Multi-Prefetcher Setups}}
\shepherd{Proba may be used as the spatial component of multi-prefetcher stacks, such as those in Neoverse V2~\cite{Arm-Hotchips}, complementing prefetchers for other access patterns. Figure~\ref{fig:hybrid-configs-ipcdram} evaluates two such configurations. In Config 1, Berti~\cite{Berti} is placed at L1 with each SMS-based prefetcher at L2. On SPECspeed 2017, Proba achieves a 3.8\% geomean speedup versus 3.1\% for the next-best design, PMP, while adding only 8.7\% DRAM traffic, compared with 15.1\% for Gaze, the next most traffic-efficient alternative. Config 2 uses a stride prefetcher at L1 and Pythia~\cite{Pythia} alongside each SMS-based prefetcher at L2. Proba improves performance by 1.7\% over the corresponding non-SMS stack, versus 1.3\% for SMS and Gaze, while again incurring the lowest DRAM traffic overhead of 5.0\%. Thus, even as the underlying prefetcher stack strengthens, Proba provides the best performance-traffic trade-off.}

\subsection{Parameter Sensitivity}
\label{ssec:sensitivity}

\shepherd{\textbf{Jail Table Size: }
Using a substantially larger JT has negligible impact on performance, implying very few false positives and negatives. Across SPECspeed 2017, increasing each JT bitmap from 4096 to 16384 bits improves geomean speedup only from 6.6\% to 6.7\%. Reducing each JT bitmap to 1024 bits increases collision-induced false positives/negatives for some workloads, causing region splitting and less accurate generation tracking. Speedup decreases from 26.8\% to 23.7\% for \texttt{cactuBSSN}, and from 19.1\% to 17.7\% for \texttt{roms}.}

\shepherd{\textbf{AGT Sampling Rate: }
We swept the AGT sampling rate from 1\% to 50\% on SPECspeed 2017. At a 50\% sampling rate, only 7.5\% of regions reach EoG. Reducing the sampling rate to 1\% increases the EoG percentage to 90.8\%, but also increases DRAM traffic overhead to 10.2\% because fewer regions are captured by the AGT. Compared with the 5\% default, 50\% reduces speedup by 0.2 percentage points, while 1\% achieves similar speedup.}

\shepherd{\textbf{Accuracy Threshold: }
Proba is surprisingly insensitive to the accuracy threshold (\textit{AccThr}) over a wide range of settings. In SPECspeed 2017, the geomean speedup varies only slightly, from 6.59\% at \textit{AccThr}=80 to 6.66\%, 6.66\%, and 6.59\% at \textit{AccThr}=40, 60, and 20, respectively. However, a higher threshold is beneficial for memory-intensive workloads because it suppresses unnecessary traffic. For example, with \textit{AccThr}=80, normalized DRAM traffic overhead for \texttt{mcf} falls to 16.9\%, compared with 27.0\%, 25.3\%, and 23.7\% at \textit{AccThr}=20, 40, and 60, respectively. We attribute this insensitivity to marginal accuracy  (\Cref{ssec:marginalpht}) when measured over multiple PHTs,  which causes filtering out of low-quality patterns across a broad range of thresholds: we observed that single-PHT variants of Proba were more sensitive to \textit{AccThr}.}

\shepherdtwo{\textbf{Number of States per PHT Entry: } We also varied the number of transition state bits (\Cref{fig:acc-cov-values}) from 1 to 4, corresponding to 2 to 16 probability states. This also has negligible performance impact across SPECspeed 2017: 1-bit, 2-bit, 3-bit and 4-bit Proba all give 6.6\% geomean speedup, though 1-bit and 4-bit increase DRAM traffic overhead to 9.9\% and 9.0\% respectively. Proba is largely insensitive to the granularity of the probability as long as it can move between several states.}

\shepherdtwo{\textbf{Larger PHTs: } Proba is insensitive to PHT sizes within reasonable ranges. Halving the number of entries per PHT to 128 reduces SPECspeed 2017's speedup from 6.59\% to 6.57\%, and increasing to 512 increases speedup to 6.64\%. DRAM traffic is marginally more sensitive, reducing to 8.1\% overhead at 128 entries (compared with 8.45\% default) and increasing to 8.71\% with 512 entries.}

\section{Related Work}
Prefetching has long been an extensively studied approach to tolerate high memory latency, with a wide range of algorithms~\cite{1197_Babak,1198_Mittal}. We discuss the most relevant below. %

\textbf{Spatial Memory Streaming Prefetchers:} Spatial Memory Streaming (SMS)~\cite{SMS} extracts spatial correlations across memory regions using bit patterns indexed by the \texttt{PC+Offset} trigger feature. Extensions of SMS~\cite{ferdman2011spatial} use rotated bit patterns to remove the offset from the trigger feature, reducing storage requirements. Later work on SMS~\cite{somogyi2011spatial} provides three static variants: SMS-AND (favoring accuracy), SMS-OR (favoring coverage), and Saturating Counters (balancing accuracy and coverage). SMS-AND and SMS-OR store two histories and apply bitwise AND or OR to issue prefetches. More recently, Bingo~\cite{Bingo} extended SMS by leveraging short- and long-term histories, inspired by the TAGE branch predictor~\cite{10.1145/2155620.2155635}. Unlike SMS, which relies on one type of trigger feature, Bingo unifies \texttt{PC+Offset} and \texttt{PC+Address} trigger features in one prediction table. DSPatch~\cite{DSPatch} improves upon SMS by trading off accuracy and coverage at runtime. It stores two predictions: one biased toward accuracy and one toward coverage. DSPatch dynamically selects between them based on bandwidth utilization. Gaze~\cite{Gaze} uses the temporal \texttt{Offset+Offset} trigger feature to index its prediction tables. PARS~\cite{10745807} supports multiple region sizes. PMP~\cite{PMP} uses saturating-counter updates and maintains two PHT tables: one based on \texttt{PC-Only} and one on \texttt{Offset-Only}.

\textbf{Aggression Control:}  Feedback-directed \cite{4147648} prefetching trades off accuracy and coverage by estimating prefetch accuracy, timeliness, and prefetch-induced cache pollution to adjust aggression. More recent work \cite{8980306} uses Perceptron-based Prefetch Filtering (PPF) to allow the underlying prefetcher to achieve high coverage while filtering out inaccurate prefetches. Similar to our work, DSPatch~\cite{DSPatch} uses table-based accuracy estimates to guide accuracy-coverage seeking decisions and control aggression.

\textbf{Other classes of prefetchers:}  Beyond spatial memory streaming, other classes of prefetchers exist, some of which are often used alongside SMS. Stride prefetchers track linear patterns with low complexity but limited coverage~\cite{381947}. More advanced schemes such as Best-Offset Prefetching (BOP)~\cite{BOP} and Berti~\cite{Berti} track more generalized past offsets. Signature Path Prefetcher (SPP)~\cite{SPP} captures correlation across multiple load PCs. Address-delta correlation can also be implemented in global history buffers~\cite{GHB}. Neural-network-based prefetchers~\cite{Hierarchical,NewForm} attempt to generalize complex patterns at higher cost. Temporal prefetchers~\cite{Triangel,Triage-MICRO19,TriageISR-ToC22,MISB-ISCA19} exploit recurrence in access sequences, while runahead schemes~\cite{DVR,391,1003_hashemi,1162_naithani,1263_Naithani,SVR} explore transient pre-execution.

\section{Conclusion}
This paper has presented \name{}, a probabilistic update mechanism for SMS-based prefetchers. By adapting spatial pattern according to accuracy-aware probabilities, \name{} preserves long-term access behavior while avoiding pattern corruption, duplicated storage, and coarse-grained update policies in prior designs such as DSPatch and PMP. It further incorporates heterogeneous trigger features from recent SMS-based prefetchers, including Gaze, while controlling their aggressiveness through marginal accuracy. Across our evaluation, \name{} consistently outperforms state-of-the-art SMS-based prefetchers, achieving higher speedups with substantially lower DRAM traffic. These results show that probabilistic pattern updates, combined with marginal-accuracy-based aggression control and multi-feature PHTs, provide an effective and practical direction for spatial prefetching in next-generation processors.

\newpage
\section{Artifact Appendix}

\subsection{Abstract}

This artifact contains the source code and simulation framework used to
evaluate Proba. The repository contains:
\begin{itemize}
    \item Source code for Proba.
    \item Source code for SMS~\cite{SMS},
          Bingo~\cite{bakhshalipourBingo2019},
          DSPatch~\cite{DSPatch},
          PMP~\cite{PMP}, and
          Gaze~\cite{Gaze}.
    \item The modified ChampSim framework and system configuration used
          to evaluate all presented prefetchers.
    \item Scripts to download SPECspeed 2017 traces, run simulations,
          and collect results.
\end{itemize}

This artifact reproduces the SPECspeed 2017 performance, DRAM traffic,
coverage, and accuracy results presented in the paper. In particular,
it reproduces the data reported in Figures~12--14.

\subsection{Artifact check-list (meta-information)}

{\small
\begin{itemize}
  \item \textbf{Program:} ChampSim
  \item \textbf{Data set:} SPECspeed 2017, downloaded using the provided scripts
  \item \textbf{Metrics:} Speedup, normalized DRAM traffic, LLC coverage,
        and overall prefetch accuracy
  \item \textbf{Output:} Tables containing per-benchmark and aggregate
        metrics for each prefetcher
  \item \textbf{Experiments:} Baseline and Proba. Optional: SMS, Bingo,
        DSPatch, PMP, and Gaze
  \item \textbf{How much disk space is required?} Approximately 40 GB
  \item \textbf{How much time is needed to prepare the workflow?}
        Approximately 1 hour to download the workload traces
  \item \textbf{How much time is needed to complete the experiments?}
        Approximately 6 hours to reproduce the main results
        (baseline and Proba). Optionally, approximately 21 hours are
        required to reproduce the results for the baseline and all
        evaluated prefetchers. These times were measured on an Intel
        Xeon Platinum 8168 machine using 40 cores.
  \item \textbf{Publicly available?} Yes
  \item \textbf{Code license:} Apache License 2.0
  \item \textbf{Archived (provide DOI):} \url{https://doi.org/10.5281/zenodo.21859952}
\end{itemize}
}

\subsection{Description}

\subsubsection{How to access}

The source code is publicly available on GitHub at
\url{https://github.com/justinhuang666666/Proba_Artifact} and on Zenodo at \url{https://doi.org/10.5281/zenodo.21859952}.

\subsubsection{Software dependencies}

Installing and running this artifact requires the following software
dependencies:
\begin{itemize}
    \item Python 3.9 or later
    \item GCC/G++ 11
    \item GNU Make
    \item Git
\end{itemize}

The artifact was tested with the following system configuration:
\begin{itemize}
    \item Ubuntu 22.04.5 LTS
    \item Linux kernel 5.15.0
    \item GCC/G++ 11.4.0
    \item Python 3.10.12
\end{itemize}

GCC/G++ 11 is required to compile the artifact. Newer compiler versions are not currently supported. The required compiler and other dependencies can be installed automatically using the setup script.

\subsubsection{Data sets}

This artifact contains a script to download the SPECspeed 2017 traces
used in DPC3~\cite{champsim_traces_speccpu_2025}, hosted at
\url{https://dpc3.compas.cs.stonybrook.edu/champsim-traces/speccpu/}.

By default, the downloaded SPECspeed 2017 traces are stored in:

\begin{artifactbox}
traces/spec2017/
\end{artifactbox}

This artifact can also be modified to reproduce the results for the
other evaluated benchmark suites. The SPEC CPU2006 traces are hosted
in the same archive as the SPECspeed 2017 traces. The Ligra traces are
available at \url{https://zenodo.org/records/14267978}, while the
CloudSuite traces are available through the Second Cache Replacement
Championship (CRC-2) at \url{https://crc2.ece.tamu.edu/}.

\subsection{Installation}

Use the following commands to clone the repository, build the ChampSim dependencies, and download the required
traces:

\begin{artifactbox}
# Clone the repository and initialize the artifact:
git clone
cd Proba_Artifact
./scripts/setup.sh

# Download the SPECspeed 2017 traces
python3 scripts/download_spec2017_traces.py
\end{artifactbox}

\subsection{Experiment workflow}

In the commands below, replace \texttt{N} with the number of simulations
to execute in parallel. For example, use \texttt{-j 40} to execute up to
40 simulations concurrently.

The evaluation baseline contains a degree-3 stride prefetcher at L1D
and no prefetcher at L2C. Therefore, the \texttt{no} configuration
denotes no \emph{L2C} prefetching, rather than no prefetching anywhere
in the cache hierarchy.

\subsubsection{Reproducing Proba}

\begin{artifactbox}
# Run the baseline and Proba configurations
python3 scripts/run_experiments.py no -j N
python3 scripts/run_experiments.py proba -j N
\end{artifactbox}

Each configuration takes approximately 3 hours on an Intel Xeon
Platinum 8168 system when running 40 simulations in parallel. Systems
with fewer available cores should use a smaller value of \texttt{N},
with a corresponding increase in total execution time.

Each simulation produces a text file containing the simulator log and
a JSON file containing the statistics used by the result-reporting
script:

\begin{artifactbox}
results/spec2017/<prefetcher>/<benchmark>/<trace>.txt
results/spec2017/<prefetcher>/<benchmark>/<trace>.json
\end{artifactbox}

\subsubsection{Optional: reproducing all evaluated prefetchers}

\begin{artifactbox}
# Run the baseline and all evaluated prefetchers
for pref in no sms bingo dspatch pmp gaze proba; do
    python3 scripts/run_experiments.py "\$pref" -j N
done
\end{artifactbox}

With \texttt{N=40}, the complete experiment takes approximately 21
hours on an Intel Xeon Platinum 8168 system.

\subsection{Evaluation and expected results}

The \path{scripts/print_results.py} script produces tables containing
the per-benchmark and aggregate results for:
\begin{itemize}
    \item IPC speedup over the baseline;
    \item normalized DRAM traffic;
    \item LLC prefetch coverage; and
    \item overall prefetch accuracy across both L2C and LLC.
\end{itemize}

Accuracy is measured across both cache levels to ensure a fair comparison
with prefetchers that employ multilevel prefetching.

\subsubsection{Reproducing Proba}

After both the baseline and Proba simulations have completed, run:

\begin{artifactbox}
# Print the Proba results
python3 scripts/print_results.py proba
\end{artifactbox}

The aggregate results should be:

\begin{resultbox}
=======================================================
> Overall summary
=======================================================
Prefetcher    Speedup   DRAM      LLC cov.  Accuracy
----------    --------  --------  --------  --------
proba         1.0659    1.0845    0.5088    0.8068
\end{resultbox}

\subsubsection{Optional: reproducing all evaluated prefetchers}

After all simulations have completed, run:

\begin{artifactbox}
# Print all prefetchers results
python3 scripts/print_results.py sms bingo dspatch pmp gaze proba
\end{artifactbox}

The aggregate results should be:

\begin{resultbox}
=======================================================
> Overall summary
=======================================================
Prefetcher    Speedup   DRAM      LLC cov.  Accuracy
----------    --------  --------  --------  --------
sms           1.0456    1.1316    0.4622    0.7699
bingo         1.0446    1.1638    0.4808    0.7516
dspatch       1.0544    1.2772    0.4247    0.7042
pmp           1.0559    1.2806    0.5751    0.6629
gaze          1.0497    1.1236    0.5372    0.6285
proba         1.0659    1.0845    0.5088    0.8068
\end{resultbox}

\subsection{Experiment customization}

Proba is implemented in \path{prefetcher/proba/}. You can change the parameter of Proba in \path{prefetcher/proba/proba.h}. The simulated processor and cache parameters can be changed in \path{params/baseline.json}. The experiment scripts under
\path{scripts/} can be modified to evaluate additional workloads,
configurations, and prefetchers.

\subsection{Methodology}

Submission, reviewing, and badging methodology:

\begin{itemize}
  \item
  \url{https://www.acm.org/publications/policies/artifact-review-and-badging-current}
  \item
  \url{https://cTuning.org/ae}
\end{itemize}

\bibliographystyle{acm}

{\balance
\bibliography{refs}
}

\end{document}